\documentclass[
12pt]{article}
\usepackage{color}
\usepackage{graphicx}
 \usepackage{amsmath,amssymb,array,calc,rotating,epsfig,psfrag}  
\usepackage[nosort]{cite}

\definecolor{darkgreen}{rgb}{0,0.55,0}

\newcommand{\bea}{\begin{eqnarray}}
\newcommand{\eea}{\end{eqnarray}}
\newcommand{\be}{\begin{equation}}
\newcommand{\ee}{\end{equation}}

\def\revise#1       {\raisebox{-0em}{\rule{3pt}{1em}}%
                     \marginpar{\raisebox{.5em}{\vrule width3pt\  
                     \vrule width0pt height 0pt depth0.5em  
                     \hbox to 0cm{\hspace{0cm}{%
                     \parbox[t]{4em}{\raggedright\footnotesize{#1}}}\hss}}}}

\def\sqr#1#2{{\vcenter{\vbox{\hrule height.#2pt    
 \hbox{\vrule width.#2pt height#1pt \kern#1pt  
 \vrule width.#2pt}\hrule height.#2pt}}}}

\catcode`\@=12

\begin{document}

\makeatletter \@addtoreset{equation}{section} \makeatother  
\renewcommand{\theequation}{\thesection.\arabic{equation}}  

\renewcommand\baselinestretch{1.25}
\setlength{\paperheight}{11in}
\setlength{\paperwidth}{9.5in}
\setlength{\textwidth}{\paperwidth-2.4in}     \hoffset= -.3in   
\setlength{\textheight}{\paperheight-2.4in}   \topmargin= -.6in 

\begin{titlepage}  

 \hbox to \hsize{{\tt hep-th/yymmnnn}\hss
\vtop{
\hbox{MCTP-08-48}
\hbox{NSF-KITP-08-66}
}}

\vskip 3cm  
 
 \begin{center}
{\bf \Large The Space-Cone Gauge, Lorentz Invariance and On-Shell Recursion for One-Loop Yang-Mills amplitudes }  
\vskip 1cm  
{\large }  
\vskip 1cm

{\large  Diana Vaman${}^1$, York-Peng Yao${}^2$  }
 
\end{center}
\vskip .2cm   
\centerline{\it ${}^1$ Department of Physics, The University of Virginia}
\centerline{\it Charlottesville, VA, 22904}
{\centerline {\it ${}^1$ KITP, University of California}}
{\centerline {\it Santa Barbara, CA, 93106}}
\centerline{\it ${}^2$ Michigan Center for Theoretical  
Physics}  
\centerline{ \it Randall Laboratory of Physics, The University of  
Michigan}  
\centerline{\it Ann Arbor, MI 48109-1120}  
 
\vspace{1cm}  
 
\begin{abstract}  
 Recursion relations are succinctly obtained for
$(++\dots +)$ and $(-++\dots +)$ amplitudes in the context
of the space-cone gauge in QCD.  We rely on the helicity symmetry of
the problems to dictate our choices of reference twistors
and the momentum shifts to complexify the amplitudes.  Of great
importance is the power of gauge Lorentz invariance, which
is enough to determine the soft factors in the latter cases.

\end{abstract}  
 
\end{titlepage}

 

\section*{}

\noindent
\section{Introduction}

An extremely efficient way to calculate multi-particle amplitudes is to
use recurrence relations, if in fact they can be devised.  Then, one can
recycle the results obtained earlier for amplitudes with smaller number
of particles to generate those with larger numbers.  In other words, the aim is
to factorize the amplitudes into sums of two groups through the exchange of
some states, where both groups correspond to physical on-shell amplitudes
which have already been constructed.  The feasibility to accomplish
this relies on analytic continuation of the amplitudes into complex momenta,
so as to be able to set the sub-amplitudes on-shell.  Obviously, one
has to have sufficient understanding and some handles on the analytic
properties of the amplitudes to push this program through.  For tree
amplitudes in QCD, because they are made of rational functions of
holomorphic $\langle ,\rangle $ and anti-holomorphic $[ \ , ] $ products, the analyticity
is simple.  Britto, Cachazo, Feng and Witten (BCFW) \cite{Britto:2005fq} have exploited it to obtain their celebrated recurrence
relations, in which the intermediate exchanges are one-particle states.
At the one-loop level, there are other amplitudes, such as those in which all or all but one
of the gluons share the same helicity, which are also rational functions
of $\langle , \rangle$ and $[ \ ,  ]$ products.  Their evaluation
requires extra physical input, such as collinearity of massless particles.
These considerations have been discussed by
\cite{Bern:1993qk}. The case when all gluons but have the same helicity was
solved by Mahlon \cite{Mahlon:1993si}, who obtained a somewhat complicated off-shell recursion relation, extending the off-shell tree level recursion relations of Berends and Giele \cite{Berends:1987me}.

In \cite{vy}, among other results, we re-derived BCFW relations
by using QCD in the space-cone gauge.  The motivation for us to
seek guidance from Lagrangian field theory is definitely not to regress
to the cumbersome perturbative diagrammatic computation, but to better
understand the make-up of the Feynman rules and therefore to explore
various ways to continue into the complex plane.

One distinct feature in the space-cone gauge of Chalmers and Siegel \cite{Chalmers:1998jb} is that QCD is like a field
theory with two scalars, corresponding to the two polarizations of a
gluon.  The redundant degrees of freedom have been eliminated to
produce non-local vertices.  The structure for a process is better
organized, because of fewer terms in each vertex and because the
number of diagrams is highly reduced.  It behooves to preserve these
virtues when we analytically continue.  In other words, let the continuation
be carried out by shifting some of the external momenta parameterized
by a variable $z$, then we shall restrict to those shifts such that
the vertices and the polarizations are not affected.  The dependence on
$z$ will come only from the scalar propagators \cite{vy}.  If we call the continued
amplitude $A(z)$, then $A(0)$ is the original physical amplitude.  A
recurrence relation follows after we complexify $z$ and   evaluate an
integral
\be
I=\oint dz {A(z)\over z},\label{(1.1)}
\ee
through a closed contour at infinity.  Because of the way we introduce $z$ dependence, as long as
there are enough numbers of scalar propagators, we shall never
pick up any surface term due to $A(z\to \infty )\ne 0.$  The recurrence
relation obtained in this fashion is on-shell, as it is made of previously
obtained lower point
amplitudes.  There is no unknown asymptotic value of $A(z)$ and
$I=0$ (see also \cite{vy1}).

One of our objectives in
this present article is to continue investigating this aspect and to apply it
to derive on-shell recurrence relations for the rational one-loop gluon amplitudes. These on-shell recurrence relations by themselves do not represent a new result, but we can offer a guiding principle for the essential step of performing the analytic continuation of the loop amplitudes. In addition, we are able to derive the soft factors which were conjectured by Bern, Dixon and Kosower (BDK) \cite{Bern:2005hs}, and which are a crucial ingredient in writing the recurrence relation, each time there is a double $z$ pole in the analytically continued amplitude. It has to be said that these double poles cannot be avoided, as they arise from the factorization of the one-loop amplitude into one-particle-reducible graphs with a one-loop $(+++)$ (or $(---)$) vertex. Our derivation of the soft factors is based on making a certain analytic continuation, with an arbitrary twistor-dependence, using the space-cone gauge to select this analytic continuation, and requiring that the end result is independent on the choice made for that arbitrary twistor. To some extent this is tantamount to requiring that the amplitude is Lorentz invariant. However, there is more to it than just Lorentz invariance, as the analytic continuation we perform is intimately related to the gauge choice. We shall refer to this as reference-twistor invariance.
The present considerations may be extended to the other one-loop amplitudes, but this falls outside the scope of this article.
For recent results on on-shell methods in perturbative QCD see \cite{Schwinn:2007ee}, \cite{Bern:2007dw}, \cite{Berger:2008sj}, and for an extension of on-shell recursions to higher-dimensions and of the space-cone gauge for gravity see \cite{ArkaniHamed:2008yf}.

The plan of this article is as follows.  In the next section,
we shall introduce the space-cone gauge, which carries
reference twistors $|+\rangle$ and $|-]$.  The resultant
Lagrangian is displayed, in which of note is that the
derivative component $\bar \partial $ does not appear
in the interaction.  This important feature dictates
the allowable choices of shifts in momenta we shall make to
complexify  an amplitude, which we discuss in details
in Section 3.  An immediate consequence is that the
analytically continued amplitude  $A(z)$ under consideration
vanishes asymptotically $z\to \infty$, which is a key
element to obtain recursion relations.  It follows in Section 4
that for the $(++\dots +)$ amplitudes, the designated
shifts should be holomorphic, involving three momenta
and having only $|+\rangle$.  Recursion relations emerge easily.
Then, we turn our attention to $(-++\dots +)$ amplitudes.
The shifts are still on a set of three momenta, but now depend on
$|-]$ instead.  Here, we shall show how the recursion
relations are organized and the soft factors are determined
due to the demand that the physical amplitudes should
be reference-twistor independent.  A few examples are
given first and then results for the general case are inferred.
A brief Section 5 will give some concluding remarks.


\noindent

\section{The space-cone gauge Yang-Mills Lagrangian}

We shall use upper case letters to denote
tensors carrying Lorentz indices and
the corresponding lower case letters for their components.
Thus, for  a four-vector $V^\mu$, we have

$$ v^{+}={1\over \sqrt2}(V^{3}+V^{0}), \ \ \  
v^{-}={1\over \sqrt2}(V^{3}-V^{0}),$$
$$ v={1\over \sqrt2}(V^{1}+iV^{2}), \ \ \  \bar v={1\over \sqrt2}(V^{1}-iV^{2}).$$
The scalar product of two four-vectors is then

$$V_\mu U^\mu=v^+u^-+v^-u^++v\bar u+\bar vu.$$

An equivalent decomposition can be introduced via a bi-spinor (twistor-product) basis:
  $$V=v^+ |+\rangle[+| + v^-|-\rangle [v| + v|-\rangle [+|+
\bar v |+\rangle [-|$$
where the reference twistors have been normalized to one: $\langle +-\rangle=
[-+]=1$. Also, any null vector can be written as
\be
V\cdot V=0 \Leftrightarrow V=|V\rangle[V|.
\ee

The space-cone gauge \cite{Chalmers:1998jb} is specified by a null  vector, which is defined by a twistor product
\be
\eta=|+\rangle[-|,
\ee
such that for the gauge field $A_\mu $ a
condition is imposed
\be
\eta\cdot A=0,
\ee
or
\be
a=0.
\ee
After re-expressing the dependent component
$\bar a$ by $a^\pm $, we have the Yang-Mills Lagrangian
\be
L=a_a^+\partial_\mu\partial^\mu a^-_a -gf_{abc}
({\partial ^-\over \partial} a^+_a)a^+_b \partial a^-_c
-gf_{abc}({\partial ^+\over \partial} a^-_a)a^-_b \partial a^+_c
+{g^2\over 2}f_{abc}a^+_a\partial a^-_b{1\over \partial ^2}
f_{a'b'c}a^-_{a'}\partial a^+_{b'}.
\ee
$f_{abc}$ are the structure constants of the gauge group,
which will be chosen as $SU(N)$.
The coupling constant $g$ used here is twice of that in the
standard quark gluon coupling $g_{QCD} \bar \psi \gamma _\mu {\lambda _a\over 2}\psi A^\mu_a$ with
$Tr( {\lambda _a\over 2}{\lambda _b\over 2})={1\over 2} \delta _{ab}.$

Note that the derivative couplings are independent of $\bar \partial$. After Fourier-transforming to momentum space, the vertices will be independent of the
four-momentum component $\bar p$. This will be relevant in all our future considerations.

\noindent

\section{Possible Analytic Continuations}

One way to continue an amplitude into complex external momenta
is to shift some of them with a parameter $z$, about which we would like
to make a comment.  In \cite{vy}, the on-shell recurrence relations were
obtained as a consequence of the largest time equation, in which some
time-ordered (or null-ordered) step functions appear.  
$z$ is the variable of integration along
the  $\underline  {real}$ axis to represent such step functions.
\be
\theta (\eta \cdot (x_1-x_2))={1\over 2\pi i}\int _{-\infty}^\infty dz
{e^{iz\eta \cdot (x_1-x_2)}\over z-i\epsilon}.  \label{(2.1)}
\ee
The ensuing complexification of $z$ is used  as a means to evaluate
subsequent integrals by using the residue theorem in (\ref{(1.1)}).  We shall abide
by this understanding and the parameter $z$ that we shall introduce
in the following shifts is assumed to be real  {\it ab initio.} At this stage
$\eta$ need not be identified with the null space-cone gauge fixing vector, but this can be a winning strategy once the analytic continuation in $z$ has been performed.

There are several
restrictions we need to observe, however.  Let the set of momenta to be
shifted among the $n$ external momenta
be  $\{P_{i_1}, P_{i_2},\dots P_{i_m}\} ,$
$n> m\ge 2,$  which we shall relabel  as $\{Q_1,Q_2, \dots Q_m\}$ for ease
of notation.  First of all, we want the shifted momenta $\hat Q_i$ to remain
massless.  This means that we write for each one
\be
\hat Q_i= Q_i +\delta Q_i=
|q_i\rangle[q_i|+ z( |q_i\rangle[\xi_i| + |\zeta_i\rangle [q_i|)
\label{(2.2)}
\ee
where $|\zeta _i\rangle$ and $[\xi _i|$ are to be determined.  Now the
overall energy-momenta must be conserved, which gives
\be
\sum_{i=1}^m |\hat Q_i\rangle[\hat Q_i|=\sum_{i=1}^m |Q_i\rangle[Q_i|, \label{(2.3)},
\ee
or
\be
\sum_{i=1}^m(|Q_i\rangle[\xi_i|+|\zeta_i\rangle[Q_i|)=0.
\label{(2.5)}
\ee

Then we must take into account that when we evaluate the integral in (\ref{(2.1)}) we are looking for singularities in $A(z)$.  It will make the task
easier if a pole in a multi-particle channel with non-vanishing
residues corresponds to a pole in $z$.  Such a
pole $(\sum P_k)^2=0,$  where $P_k$ are some of the n external momenta,
contains products of some of the shifted momenta
\be
-2\hat Q_i \cdot \hat Q_j=\langle\hat Q_i \hat Q_j\rangle[\hat Q_j \hat Q_i].  \label{(2.7)}
\ee
Clearly, we have terms $\sim z^{0,1,2}$.  To have just a pole in $z$ when
we solve for it from $(\sum P_j)^2=0$, we must demand that the coefficients to
$z^{2}$ vanish:
\bea
&&\langle\zeta _i \zeta_j\rangle =0, \label{(2.8)}\\
&&[\xi_i \xi_j]=0. \label{(2.9)}
\eea

The implication of (\ref{(2.8)}) is immediate, namely some of the
$|\zeta_i\rangle$ are null  and that those which are not must be aligned.  The
same is said for $[\xi_i|, $ because of (\ref{(2.9)}).    
Hence, we can write  
\be
| \zeta_i\rangle=c_i |\zeta\rangle, \ \ \ \  |\xi_i]=d_i|\xi],\label{(2.11)}
\ee
for those $Q_i$ and $Q_j$ which can be grouped together into at least
one of the possible channels that form single particle poles.  

At this point, we may have some $Q_i$'s and $Q_j$'s which cannot be
grouped together onto the same side of any channel $\sum P_k$.  As a consequence, we may
have several pairs of $|\zeta\rangle $ and $[\xi|$, but in the space-cone
gauge, the most natural thing to do is to collapse them into just one pair.

There are other conditions on $c_i, \ d_i, \  |\zeta\rangle$ and $|\xi]$ which follow from the requirement that the relevant polarization vectors and vertices
should not have any $z$ dependence. We want to enforce these conditions in order for the analytically continued amplitude $A(z)$ to be well behaved at
infinity.
 
As said, it is natural to make the identification
\be
|\zeta \rangle=|+\rangle , \qquad |\xi]=|-]. \label{(2.16)}
\ee
For those $Q_j$ with positive helicity, the external line factors are
\be
\epsilon ^+(Q_j)={[-Q_j]\over \langle+Q_j\rangle},\label{(2.19)}
\ee
and those $Q_k$ with negative helicity
\be
\epsilon ^-(Q_k)={\langle+Q_k\rangle\over [-Q_k]}.\label{(2.20)}
\ee
We note that because of the identification in (\ref{(2.16)}), we have
\be
\epsilon ^+(\hat Q_j)=\epsilon ^+(Q_j), \ \ \ \epsilon ^-(\hat Q_k)=\epsilon ^-(Q_k),
 \ \ \   \hat q_i=q_i.\label{(2.21)}
\ee
The next important observation is that the vertices in this gauge depend only
on $(k^\pm, k)$ and $(p_i^\pm, p_i)$ but not on $\bar k$ or $\bar p_i$,
where $K_\mu $ is the loop momentum.  After performing the loop integration, we
shall have only tensors made of $(p_i^\pm, p_i)$. Therefore, our choice
of $c_i$ and $d_i$ is to make
\be
\hat q_i^\pm= q_i^\pm,\label{(2.22)}
\ee
whichever appear in the amplitude.  Whence, all the $z$ dependence
is confined in the denominators of some scalar integrals of  Feynman parameters.

To summarize, we find that the shifts of (\ref{(2.2)}) have to satisfy
(\ref{(2.11)}, \ref{(2.22)}) and (\ref{(2.5)}).  
The latter is written as
\be
\bigg(\sum_{i=1}^m d_i|Q_i\rangle\bigg)[-|\,+\,|+\rangle \bigg(\sum_{i=1}^m c_i[Q_i|\bigg)=0. \label{(2.23)}
\ee
We shall consider $m=2,3$ in the following.

For $m=2$, we first look at  $Q_1$ and $Q_2$ both of which come with positive helicity
gluons.  Let us first consider the consequence of $\hat q_1^-=q_1^-$ and
$\hat q_2^-=q_2^-$, which yield $d_1\langle+Q_1\rangle =d_2\langle+Q_2\rangle =0,$ or
\be
d_1=d_2=0, \label{(2.24)}
\ee
because otherwise we shall need $\langle+Q_1\rangle =0,$ and/or $\langle+Q_2\rangle =0.$  These
conditions will give rise to very badly behaved polarizations  $\epsilon^+(Q_1)$
and/or $\epsilon^+(Q_2)$.

With the conditions of (\ref{(2.24)}), (\ref{(2.23)})
is now $c_1|Q_1]+c_2|Q_2]=0,$ which in turn gives
\be
c_1=c_2=0 \label{(2.25)}
\ee
meaning no shift.  We are then led to the conclusion that we cannot analytically continue
into complex momenta by shifting only two of the momenta, if both of
their gluons have the same helicity.  It then follows trivially that
for amplitudes with all positive or negative helicity, one has to shift
more than two momenta  for continuation.

Now we turn to the case when $Q_1$ comes with positive helicity but
$Q_2$ with negative helicity.  Our requirement $\hat q_{1,2}^\pm=q_{1,2}^\pm$
yields
\be
c_1[Q_1-]=0, \ \ \ \  c_2[Q_2-]=0, \label{(2.26)}
\ee
and
\be
d_1\langle+Q_1\rangle =0, \ \ \ \  d_2\langle+Q_2\rangle =0.\label{(2.27)}
\ee
We pick
\be
d_1=0  \ \ \ \  {\rm and} \ \ \ \ c_2=0 \label{(2.28)}
\ee
so that $\epsilon^+(Q_1)$ and $\epsilon^-(Q_2)$ behave properly.  In
order not to have a trivial shift, we must have also $[Q_1-]=0$ and
$\langle Q_2+\rangle=0,$ or
\be
|Q_1]=|-], \ \ \ \ {\rm and}  \ \ \ \ |Q_2\rangle =|+\rangle. \label{(2.29)}
\ee
An additional condition is from (\ref{(2.23)}), which is
$c_1+d_2=0$.  Because $z$ is arbitrary, we can just scale them to
\be
c_1=-d_2=1.\label{(2.30)}
\ee
This is the standard shift used by BCFW \cite{Britto:2005fq}.

Now we look at $m=3$.  Since one-loop rational gluon amplitudes are our focus, we first look at the case when all external gluons have the same helicity. To be specific, we consider a one-loop amplitude with all external gluons having
positive helicities.  At the one-loop level, the
space-cone gauge will not produce vertices which depend on $p_j^+$ or $q_i^+$.  Then
the demand $\hat q^-_{1,2,3}=q^-_{1,2,3}$ gives
\be
d_1=d_2=d_3=0.\label{(2.31)}
\ee
Then all the $c_{1,2,3}$ should not
vanish, because we want to have genuine shifts involving three of
the momenta.  From (\ref{(2.23)}) we find
\be
c_1|Q_1]+c_2|Q_2]+c_3|Q_3]=0.  \label{(2.32)}
\ee
The twistors live in a two-component space and we are looking
for a linear relation among three of them.  This is provided by
the Schouten identity, with a solution
\be
c_1=[Q_2Q_3], \ \ \  c_2=[Q_3Q_1], \ \ \ c_3=[Q_1Q_2].  \label{(2.33)}
\ee
These shifts were used for a different purpose by Risager in \cite{Risager:2005vk}.

Of course, one may not want to choose to work in the space-cone gauge, or may
not want to identify $|\xi]=|-]$ and $|\zeta\rangle =|+\rangle $.  There are many other shift
possibilities.  For example, one may take $d_1\ne 0,$ but $d_2=d_3=0$,
which comes with $c_1=0,$ but $c_2\ne 0,$ and $c_3\ne 0$.  Then
\be
|Q_1\rangle d_1[\xi|\,+\,|\zeta\rangle (c_2[Q_2|+c_3[Q_3|)=0, \ee
which up to some inconsequential multiplicative factor yields
\be|\zeta\rangle =|Q_1\rangle,
\ee
and
\be
|\xi]=c_2|Q_2]+c_3|Q_3] 
\ee
after setting $d_1=-1$.  These are the shifts used in \cite{Risager:2005vk}, if we
relabel $Q_1, \ Q_2, \ Q_3$ respectively as $j, \ l, \ n$, for
a particular choice of $c_2$ and $c_3$.  The downside of working outside the space-cone gauge is that we do not know a priori if these shifts are such that
the analytically continued amplitude $A(z)$ vanishes as $z\to\infty$. This means that there might be a boundary contribution from evaluating the  contour integral $\oint \frac{dz}{z} A(z)$.

For a one-loop amplitude of the type $(-+\dots+)$, the space-cone gauge Feynman diagrams are constructed from trivalent vertices only. We are also making the identification $|+\rangle\equiv |Q_1\rangle$, where $Q_1=|Q_1\rangle [Q_1|$ is the momentum of the external negative helicity gluon. Then the only $(--+)$ vertex is the one where one of the legs is $Q_1^-$, and all other vertices are of the type $(++-)$. Moreover, the vertex times external line factor for a vertex $(Q_1^-,Q_2^+ , K^-)$ is $\epsilon^+(Q_2) k$.  This means that the vertices depend only on $q_i^+, q_i$ momentum components. Therefore, a possible triple momentum shift which leaves the vertex structure and polarizations invariant is
\bea
&&|+\rangle \equiv |Q_1\rangle, |-] = {\rm arbitrary},|\zeta\rangle=|+\rangle,
|\xi]=|-]\nonumber\\
&&c_1\neq 0, c_2=0, c_3=0, d_1=0, d_2\neq 0, d_3\neq0.
\eea
Momentum conservation, together with the requirement $\hat q_i^+ = q_i$ leads to
\bea
c_1= [Q_2 Q_3], \qquad d_2=[Q_3 -],\qquad d_3= [-Q_2].
\eea


\section{On-shell recurrence relations for one-loop rational gluon amplitudes}

Scattering amplitudes are known to have certain regions of analyticity in
complexified energy momenta.  This subject was studied in earnest in
the 1960's.  Unfortunately, being such a broad thesis, analyticity by itself
was quite limited in revealing at a deeper level the structure and the
symmetry of amplitudes.  What it lacks is the constraints of dynamics.

This pursuit may have regained some currency in recent work of  QCD helicity
amplitudes (see for example \cite{Bern:2007dw}, and the references within).  For example, for tree amplitudes, where there are only simple
poles in the intermediate channels, clever use of analytic properties leads to
the revolutionary BCFW recurrence relations.  We would like to extend such
considerations to one-loop amplitudes, specifically to the case when the amplitudes remain rational functions of the external gluon momenta. These are amplitudes which vanish at the tree level.

\subsection{One-loop same helicity gluon amplitudes}


We begin by making the observation that in the space-cone gauge
a one loop same helicity gluon amplitude is built only out of 3-point
vertices. To be specific, let us consider 1-loop amplitudes
with positive helicity external gluons. Then all the vertices will be
trivalent: $(++-)$.

First, we notice that we cannot specify the space-cone vector
$\eta=|+\rangle [-|$ completely in terms of external gluon momenta. The reason
is that the external (on-shell) gluons have all the same helicity.
Usually, $\eta$ is built
by identifying the positive chirality spinor such that
$|+\rangle = |i\rangle$,
where the $i$-th external gluon has negative helicity,
and the negative chirality spinor $|-]=|j]$,  where the $j$-th external
gluon has positive helicity. Since in our case all external gluons
have positive helicity, this means that $|+\rangle$ remains arbitrary,
and distinct from any external gluon $|k\rangle$, to avoid division by zero.
Of course, the amplitude itself must be Lorentz covariant, and independent
on our choice for $|+\rangle$.

Second, to derive the on-shell recurrence relations we can no longer
rely on making the usual shifts of the two external momenta singled out
to construct the space-cone vector $\eta$, since all gluons have the same helicity. However, we do have an alternative, as discussed in the previous section: we can select three external gluons and
shift their momenta according to
\bea
&&P_1=|1\rangle [1|\rightarrow \hat P_1=|\hat 1\rangle[1|=
(|1\rangle+z[23]|+\rangle)[1|
\nonumber\\
&&P_2=|2\rangle [2|\rightarrow \hat P_2=|\hat 2\rangle[2|=
(|2\rangle+z[31]|+\rangle)[2|\nonumber\\
&&P_3=|3\rangle [3|\rightarrow \hat P_3=|\hat 3\rangle[3|=
(|3\rangle+z[12]|+\rangle)[3|. \label{holo}
\eea
This analytic continuation leaves invariant the vertices and the external line factors of the space-cone gauge, but the
internal propagators do change, with at least one of them being affected. Then
 we infer that $A(z) \sim 1/z^n$, with $n\geq 1$ as $z\to \infty$.
Therefore, $
\oint \frac{dz}z A(z)=0 $  
when the contour is taken at infinity. In other words, there is no
boundary term to contend with.
This means that
\be
0=\sum_i \frac{Res(A)}{z}\bigg|_{z=z_i}+ A(0)
\ee
and we can factorize the all plus amplitude into lower $n$-point functions
associated with the residues of $A(z)$. Since all external gluons have same
helicity, the only set of residues comes from cutting an internal line
which isolates a tree level three-point function with two of the external
gluons on one side
(at least one of them being part of the triplet which is being shifted), and
a one loop $n-1$-point function on the other side of the cut.
So we arrive at a recurrence relation of the type
\bea
A_n^{(1)}({P_1, P_2, P_3}, P_4,\dots P_n)&=&
A^{(0)}_3({ \widehat P_1, \widehat P_2}, K)
\frac{1}{{2  P_1\cdot P_2}}A_{n-1}^{(1)}(K,{ \widehat P_3},
P_4\dots P_n)\bigg|_{z_{12}}\nonumber\\
&+&
A^{(0)}_3({ \widehat P_2, \widehat P_3}, K)
\frac{1}{{2  P_2 \cdot P_3}}A_{n-1}^{(1)}(K, P_4,
\dots P_n { \widehat P_1})\bigg|_{z_{23}}\nonumber\\&+&
A^{(0)}_3({ \widehat P_3}, P_4, K) \frac{1}{{2  P_3}
\cdot P_4}A_{n-1}^{(1)}(K,
P_5\dots P_n, { \widehat P_1, \widehat P_2})\bigg|_{z_{34}}\nonumber\\&+&
A^{(0)}_3(P_n, { \widehat P_1}, K) \frac{1}{2 P_n \cdot { P_1}}
A_{n-1}^{(1)}(K,
{ \widehat P_2, \widehat P_3}, P_4\dots P_{n-1})\bigg|_{z_{n1}}\label{rec+}
\eea   
where the superscripts $(0,1)$ indicate whether the on-shell amplitude
is tree or one-loop level and the hats denote the shifts made such that the line
cut is put on-shell.
The $z$ shifts corresponding to the four terms which appear in
the previous recurrence relation are given by
\bea
&& z_{12}=-\frac{\langle 1 2\rangle}{[23]\langle + 2\rangle +[31]\langle 1
+\rangle}\nonumber\\
&&z_{23}=-\frac{\langle 23\rangle}{[31]\langle +3\rangle+[21]\langle+2\rangle}
\nonumber\\
&&z_{34}=-\frac{\langle 43\rangle}{[12]\langle4+\rangle}, \qquad
z_{n1}=-\frac{\langle n1\rangle }{[23]\langle n+\rangle}.
\eea

For instance, the all plus one loop 5-point function can be constructed out of
one-loop 4-point functions according to (\ref{rec+}):
\bea
A_5^{(1)}({ 1^+, 2^+, 3^+}, 4^+, 5^+)&=&\frac{-iN_p}{96\pi^2}\bigg(
\frac{([51]\langle 1+\rangle+[52]\langle2+\rangle)
([31]\langle 1+\rangle +[32]\langle 2+\rangle)}{\langle 12\rangle\langle
\hat 3 4\rangle \langle 45\rangle\langle 1+\rangle\langle 2+\rangle}
\bigg|_{z_{12}}
\nonumber\\
&+&\frac{(12]\langle 2+\rangle+[13]\langle 3+\rangle)([42]\langle 2+\rangle+
[43]\langle 3+\rangle)}{\langle 23\rangle\langle\hat 15\rangle\langle 54
\rangle\langle 2+\rangle\langle 3+\rangle}\bigg|_{z_{23}}\nonumber\\
&-&\frac{([53]\langle 3+\rangle+[54]\langle4+\rangle)([23]\langle 3+\rangle +[24]\langle 4+\rangle)}{\langle 34\rangle\langle\hat 1\hat 2\rangle
\langle \hat 1 5\rangle\langle 3+\rangle\langle 4+\rangle}\bigg|_{z_{34}}
\nonumber\\
&-&\frac{([41]\langle 1+\rangle+[45]\langle 5+\rangle)([21]\langle 1+\rangle
+[25]\langle 5+\rangle)}{\langle 15\rangle\langle\hat 2\hat 3\rangle\langle
\hat 3 4\rangle\langle 1+\rangle\langle 5+\rangle}\bigg|_{z_{51}} \bigg),
\eea
where $N_p$ is the number of particles circulating in the loop.
We verified using a symbolic manipulation program that the previous sum
is independent of $|+\rangle$, and that it reproduces the expected result
\bea
A_5^{(1)}({1^+, 2^+, 3^+}, 4^+, 5^+)&=&
\frac{-iN_p}{96\pi^2}
\frac{(1234)+(1235)+(1245)+(1345)+(2345)}
{\langle 12\rangle\langle 23\rangle
\dots\langle 51\rangle},
\eea
where
\bea
(1234)=\langle 12\rangle[23]\langle34\rangle [41], \;{\rm etc}\dots
\eea

\subsection{Reference-Twistor Independence and
One-Loop $(-++\dots+)$ Amplitudes}


We have argued that the $(-++\dots +)$ amplitude, being
extended to complex momenta by the shifts given in Section 2 and denoted as
$A(z)$, vanishes as $z \to \infty$.  The physical amplitude
$A(z=0)$ can be recovered by a closed contour integral

$$\oint dz {A(z)\over z}=0,$$
which gives

$$A(z=0)=-\sum_i(z-z_i){A(z)^{s.p.}\over z}\bigg|_{z=z_i}
+\sum_j (z-z_j)^2{A(z)^{d.p.}\over z^2}\bigg|_{z=z_j},$$
where we have assumed that $A(z)$ has a set of
simple poles at $z_i$ in $A(z)^{s.p.}$ and a set of
double poles at $z_j$ in $A(z)^{d.p.}$. This is an
extended recursion relation.

Now $z_i$ and $z_j$ all depend on the reference twistors $|+\rangle$
and $|-]$.  The gauge-fixing vector $\eta=|+\rangle[-|$, if present in a
physical amplitude, would even destroy Lorentz invariance.
However, if we focus on the right hand side, we actually
avail ourselves of an opportunity to study the interplay of
reference twistors dependence of individual terms.  The
dependence must be quite special, because the sum
being the physical amplitude should not allow
it.  In other words, the pieces which depend
on $|+\rangle$ and/or $[-|$ must cancel.  For a tree amplitude,
all the terms on the right hand side are made of known
tree amplitudes of smaller number of gluons and therefore
the independence is just a check on a calculation.  For
a loop amplitude, the situation is more interesting, because
there are one particle irreducible diagrams.  They give
rise to contributions to partial amplitudes with simple
poles which are buried under those with double poles
with $z_i=z_j$.  Thus, the right hand side must contain
some factors which have not been encountered before.
We shall show that, for $(-++\dots +)$ amplitudes in
QCD, the requirement that the sum should be independent
of the gauge twistors is enough to determine these unknown 'soft
factors' uniquely.  That is to say the constraints due
to gauge invariance are so stringent that the contributions
from the irreducible diagrams can be obtained without
having to calculate them explicitly.  This is not something
one can naively take for granted.  If the irreducible graphs
had some gauge invariant components, as in QED,
we would not have been able to determine them by this
procedure.

\subsubsection{The Analytic Continuation}
 
Let the one-loop amplitude $(-++\dots+)$ be specified by the momenta and helicities of the external gluons $(1^-2^+3^+\dots n^+)$.
As argued previously, the analytic continuation of this class of rational amplitudes can be done by a triple shift of the external momenta.
We choose to shift the momenta $P_1, P_2, P_n$ by:
\bea
&&P_1\to\hat P_1=|1\rangle [1|+ z[2n] |1\rangle [-|\nonumber\\
&&P_2 \to \hat P_2=|2\rangle [2| + z[n-] |1\rangle [2|\nonumber\\
&&P_n\to \hat P_n = |n\rangle [n| + z [-2] |1\rangle [n|\label{gen_shifts}
\eea
and we shall make the choice
\be
|+\rangle\equiv |1\rangle,
\ee
while keeping $|-]$ arbitrary (recall that the space-cone gauge fixing vector is $\eta=|+\rangle [-|$). In expressing the $n$-point amplitude $A$ through an on-shell recurrence relation, we need to evaluate the residues of $A(z)/z$.
Since the shifts by which we analytically continued $A(z)$ are now dependent on the twistor $|-]$, so will the poles $z_i$ of $A(z)$. Of course, the $n$-point amplitude itself, $A$, should be independent on $|-]$. This is a statement of Lorentz invariance. However, since we are not only requiring that $A$ be independent of the choice of null gauge fixing vector $\eta$, but we are using an analytic continuation $A(z)$ which we correlated with $\eta$, we are going to refer to this as reference-twistor independence of the amplitude, or
$\eta$-independence for short.
This will turn out to be a requirement powerful enough to determine the soft factors introduced by Bern, Dixon and Kosower in \cite{Bern:2005hs}.

\noindent
\subsubsection{The Four-Point (-+++) Amplitude}

Let us first consider the scattering amplitude $(1^-2^+3^+4^+)$ at
the one-loop level.  There are three diagrams which contribute in the space-cone gauge,
two of which are one particle reducible (1PR) and one is not.  Our aim is to
construct the full amplitude by imposing Lorentz invariance (or effectively, reference-twistor invariance) on some suitably
modified reducible amplitudes.  For this purpose, we analytically
continue the amplitude by shifting some of the spinors in $P_i=|i\rangle[i|, \
i=1,2,3,4$,  which are taken to be outgoing.  These shifts produce distinct poles in the $z$-plane for the
channels we are interested in, namely at
\be
s=-(P_1+P_2)^2, \ \ \  {\rm and} \ \ \  u=-(P_1+P_4)^2.
\ee
We use the following set of  triple shifts
\bea
&&|\hat 1]=|1]+z[24]|-]\nonumber\\
&&|\hat 2\rangle=|2\rangle+z[4-]|1\rangle \nonumber\\
&&|\hat 4\rangle=|4\rangle + z[-2]|1\rangle \label{4pt_ana}
\eea
and we further identify
$
|+\rangle\equiv |1\rangle.
$
The singularities of the analytically continued amplitude
\be
A_4^{(1)}(\hat 1^-,\hat2^+,3^+,\hat4^+)=A(z)
\ee
will be located at the invariants
\bea
\hat s&=&\langle\hat 1 \hat 2\rangle[\hat 2 \hat 1]=\langle12\rangle[24][2-](z-z_s),\\
\hat u&=&\langle\hat 1 \hat 4\rangle[\hat 4 \hat 1]=\langle14\rangle[24][4-](z-z_u),
\eea
where
\bea
z_s&=&-{[21]\over [2-][[24]},\qquad z_u=-{[41]\over [4-][24]}.
\eea
Note that the physical amplitude which is $A(z=0)$, and must be Lorentz invariant, i.e. with  no dependence on  reference
spinors. Given our analytical continuation in (\ref{4pt_ana}), it is trivial
to remark that at $z=0$ there is no dependence on the twistor $|-]$ whatsoever.  On
the other hand, the extended amplitude $A(z)$ has first and
second order poles in the $z$-plane at $z_{s}$ and $z_{u}$.  If we use
Cauchy integration to pick out these individual parts, their residues
are manifestly dependent on $|-] $.    
There is a clear interpretation for the second order poles in the $z$-plane: they correspond to the 1PR Feynman graphs. That is because the 1PR diagrams contain a one-loop $(+++)$ vertex, with two of the legs on-shell. This vertex is proportional to $\frac{1}{K^2}$, where
$K$ is the momentum of the off-shell gluon. Together with the $K$-line propagator this gives an overall dependence which is $(K^2)^{-2}$, and which leads to a double pole in the $z$-plane upon analytic continuation.
There is also a simple pole underneath the double pole. This is much more subtle to see. We are simply going to infer its existence based on the observation that if the double pole factorization were the whole story, then we are in trouble. The terms which correspond to the second-order poles in $z_s$ and $z_u$ add up to an expression which depends on the choice of the reference twistor $|-]$. So, if the 4-point amplitude $A(z=0)$ can be recovered from the $z$-plane residues, there must be some other singularity, besides the the double poles at $z_s$ and $z_u$. Also, from kinematic considerations, the only singularities are at $z_s$ and $z_u$. Therefore, the additional singularities can only be first order poles at $z_s$ and $z_u$.

This motivates the conjecture
\be
A_4= A_{4s}^{1PR} F_s+ A_{4u}^{1PR} F_u \label{offshell}
\ee
where $A_{4s}^{1PR}$ and $A_{4u}^{1PR}$ are the two one-loop reducible Feynman graphs, and $F_s$ and $F_u$ are dressing factors, depending only on kinematic invariants, to be determined. (Jumping ahead, $F_s=1+f_s$, $F_u=1+f_u$, where
$f_s, f_u$ are the 'soft-factors` of BDK.)

For concreteness, let us gather a few results. The one-loop amplitude corresponding to a choice of external gluon helicities $(-+++)=(1234)$ is:
\be
A_4^{(1)}=N\frac{\langle24\rangle[24]^3}{[12]\langle23\rangle\langle34\rangle[41]}
\ee
where $N=(iN_p)/(96\pi^2)$ is a one-loop numerical normalization factor.
The reducible Feynman graphs are
\bea
&&
A_{4s}^{1PR}=N\frac{[34]^3\langle+ 3\rangle\langle+4\rangle\langle1-\rangle[2-]^2}{[+-]\langle+-\rangle s^2}\equiv \frac{
a_{4s}^{1PR}}{s^2}\\
&&
A_{4u}^{1PR}=N\frac{[32]^3\langle+3\rangle\langle+2\rangle\langle1-\rangle[4-]^2}{[+-]\langle+-\rangle u^2}\equiv \frac{
a_{4u}^{1PR}}{u^2}.
\eea
After a bit of massaging, we can rewrite them as
\bea
&&a_{4s}^{1PR}=N\frac{[23][34][42]}{[1-]} [4-]\langle14\rangle\\
&&a_{4s}^{1PR}=N\frac{[23][34][42]}{[1-]} (-[2-]\langle12\rangle).
\eea

As a check of the conjecture (\ref{offshell}) let us choose $|-]=|2]$.
Then the four-point one loop amplitude is recovered provided that
\be
F_u=\frac{s+u}{s}.
\ee
If we choose instead $|-]=|4]$, then we find that we need
\be
F_s=\frac{s+u}{u}.
\ee

We shall soon find that these choices are not arbitrary and they are dictated by reference-twistor invariance (which is a statement of Lorentz invariance) and analyticity.
The reference-twistor invariance requires that the amplitude $A_4$ should be independent on the value taken by the reference twistors $|-\rangle, |-]$.
With the concrete expression of the 1PR diagrams at hand, this is ensured
provided that
\be
F_s =\frac{f(s,u)}{u},\qquad F_u=\frac{f(s,u)}{s},\label{constraint1}
\ee
where $f(s,u)$ is a function which depends only on the kinematical invariants $s,u$.
Next, we analytically continue $A_4$ to the complex plane using (\ref{4pt_ana}).
We are going to evaluate the contour integral
$\frac{1}{2\pi i}\oint_{C_\infty}\frac{dz}{z} A_4(z)$.
Since $A_4^{(1)}$ is a rational function, the integral gets localized onto the residues. We have already argued that the integral receives no contribution from the contour which is taken at infinity. It is also transparent that $A_{4s}^{1PR}(z)=\frac{A_{4s}^{1PR(0)}}{s^2(z)}$, and similarly, the $z$-dependence of $A_{4u}^{1PR}(z)$ comes only from the shifted denominator $\frac{A_{4u}^{1PR(0)}}{u^2(z)}$. So, with trivial dressing factors,
the integral would be localized only on the second-order poles $z_s$ and $z_u$.

We can solve for $f(s,u)$ as follows. First, let us choose $|-]=|2]$. Then the four-point amplitude ought to receive a contribution only from the $z=z_u$ singularity:
\be
A_{4}^{(1)}=a_{4u}^{1PR}\frac1{\langle 14\rangle^2 [24]^2}\partial_z \frac{F_u(z)}{z}\bigg|_{z=z_u}\label{a41}
\ee
where $s(z)=s$.
On the other hand, if $|-]=|4]$, then
\be
A_{4}^{(1)}=a_{4s}^{1PR}\frac1{\langle 12\rangle^2 [24]^2}\partial_z \frac{F_s(z)}{z}\bigg|_{z=z_s}\label{a42}
\ee
where $u(z)=u$.
Identifying the right hand side of equations (\ref{a41}) and (\ref{a42}) leads to a differential equation constraint for $f(s,u)$:
\be
(u\partial_u f+f)\bigg|_{s,u=0}=(s\partial_s f + f)\bigg|_{u,s=0}
\ee
which is solved by
\be
f=s+u
\ee
and so,
\be
A_{4}^{(1)}(1^-,2^+,3^+,4^+)= \frac{s+u}u A_{4s}^{1PR} + \frac{s+u}{s} A_{4u}^{1PR}.
\ee


\subsubsection{The Five-Point (-++++) Amplitude}

We want to support our contention that the reference twistor ($\eta$)-invariance
requirement can be stringent enough to be used to determine
contributions from irreducible graphs in some cases, by giving
another example:  $(1^-2^+3^+4^+5^+)$.  
It is natural to choose $|+\rangle=|1\rangle$, while keeping $|-]$ arbitrary, for the time being.  The external momenta $P_1,P_2, P_5$ are shifted according to
\bea
&&|\hat 1]=|1]+z[25]|-],\nonumber\\
&&|\hat 2\rangle=|2\rangle+z[5-]|1\rangle,\nonumber\\
&&|\hat 5\rangle=|5\rangle+z[-2]|1\rangle, \label{(14)}
\eea
The rest of the spinors
are unchanged.  The overall momenta are conserved in view of Schouten identity
$[25][-|+[5-][2|+[-2][5|=0.$

We shall make our presentation as much as possible in
the $z$-plane, consistent with what is to be expected of a complexified amplitude
$A_5(z)$.  Then, the poles in the $z$-plane come from the vanishing
of the invariants

$$\hat s_{12}=\langle12\rangle[2\hat 1], \ \ \ \hat s_{15}=\langle15\rangle[5\hat 1],
\ \ \ \hat s_{23}=\langle\hat 2 3\rangle[32], \ \ \ \hat s_{45}=\langle4\hat 5\rangle[ 54],$$
or the vanishing of $[2\hat 1], [5\hat 1], \langle\hat 2 3\rangle, \langle4 \hat 5 \rangle$ respectively.  
They yield
$$[\hat 1 2]=[25][-2](z-z_{12}), \ \ \ [5\hat1]=[25][5-](z-z_{15}),$$
$$\langle\hat 2 3\rangle=[5-]\langle13\rangle(z-z_{23}), \ \ \ \langle4\hat 5\rangle=[-2]\langle41\rangle(z-z_{45}),$$
where
\bea
&&z_{12}=-{[12]\over [25][-2]}, \ \ \ z_{15}=-{[51]\over [25][5-]}, \nonumber\\
&&z_{23}=-{\langle23\rangle\over [5-]\langle13\rangle},\ \ \ z_{45}=-{\langle45\rangle\over [-2]\langle41\rangle}.
\label{(15)}
\eea
To understand the residue at $z_{12}$, one constructs the vector
which corresponds to the intermediate cut channel $\hat P_1 +\hat P_2 \equiv \hat K_{12}$  at that position
\be
\hat K_{12}\equiv (|1\rangle[\hat 1|+|\hat 2\rangle[2|)_{z=z_{12}}
=({[15]\over[25]}|1\rangle+|2\rangle)[2|,
\ee
after using
\be
[\hat 1|_{z=z_{12}}=\frac{[1-]}{[2-]}[2|,
\ee
and
\be
|\hat 2\rangle_{z=z_{12}}=|2\rangle+\frac{[12][5-]}{[25][2-]}|1\rangle.
\ee
Thus, its twistors are
\be
|\hat K_{12}|=[2|, \ \ \ |\hat K_{12}\rangle={[15]\over [25]}|1\rangle+|2\rangle,
\ee
which are $\eta$-independent.  With these, we calculate
\be
A_{4}^{(1)}(\hat K_{12}{}^{+} ,3{}^{+},4{}^{+},\hat 5{}^{+})|_{z=z_{12}}=
N{[\hat K_{12} 3][4\hat 5]\over \langle\hat K_{12} 3\rangle\langle 4\hat 5\rangle}|_{z=z_{12}}
=N{[25]^{2}\over\langle34\rangle^{2}},
\ee
and
$$A_{3}^{(0)}(\hat 1{}^{-},\hat 2{}^{+},-\hat K_{12}{}^{-})|_{z=z_{12}}=\frac{\langle12\rangle[2-]}{[1-]},$$
and form the partial amplitude
\bea
A_{12}  &\equiv &
A_{3}^{(0)}(\hat 1{}^{-},\hat 2{}^{+},-\hat K_{12}{}^{-})|_{z=z_{12}}
{1\over K_{12}^{2}}A_{4}^{(1)}(\hat K_{12}{}^{+} ,3{}^{+},4{}^{+},\hat 5{}^{+})|_{z=z_{12}}\nonumber\\ &
=& N\frac{[25]^2[2-]}{\langle 34\rangle^2 [1-][12]},
 \label{(16)}
\eea
where we substituted
$$K_{12}=(P_1+P_2)^2=\langle12\rangle[12].$$

In a similar fashion, we evaluate the residue of $A_5$ at
$z=z_{15}$
\bea
A_{15}& \equiv& A_{3}^{(0)}(\hat 1{}^{-},\hat 5{}^{+},-\hat K_{15}{}^{-})|_{z=z_{15}}
{1\over K_{15}^{2}}A_{4}^{(1)}(\hat K_{15}{}^{+} ,\hat 2{}^{+},3{}^{+},4{}^{+})|_{z=z_{15}}\nonumber\\ &
=& N \frac{[25]^2[5-]}{\langle 34 \rangle^2 [1-] [15]}.
\label{(17)}
\eea

\begin{figure}[h!]
\begin{center}
\includegraphics[height=3in]{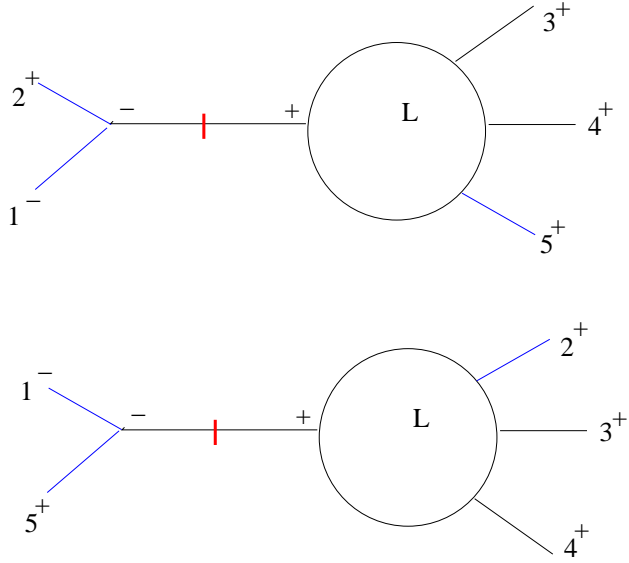}
\caption{$A_{12}$ and $A_{15}$}
\end{center}
\end{figure}

We note that the reference-twistor dependent part of $A_{12}$
cancels that of $A_{15}.$

In Figure 1 we have depicted the cut sub-amplitudes $A_{12}$ and $A_{15}$.
The line which is cut (with the cut depicted by a small vertical red line)
is placed on-shell by the shifting the momenta $P_1,P_2,P_5$. The shifts of the {\it on-shell} momenta, which are always those of the lone  negative helicity gluon $1^-$ and of the two positive helicity gluons adjacent to it, are represented by blue lines. The letter $L$ stands for a one-loop amplitude. The helicities of the on-shell gluons are specified as well:$+/ -$ for a positive/negative helicity gluon.

We would like to make a comment.  In  color-ordered
amplitudes, physical singularities appear due to the
vanishing of invariants formed with sequential momenta.
In (\ref{(16)})  and (\ref{(17)}), we see that they potentially can
have singularities in $z$ corresponding to $\langle34\rangle$ in the
denominators, which come from
$A_{4}^{(1)}(\hat K_{12}{}^{+} ,3{}^{+},4{}^{+},\hat 5{}^{+})|_{z=z_{12}}$ and
$A_{4}^{(1)}(\hat K_{15}{}^{+} ,\hat 2{}^{+},3{}^{+},4{}^{+})|_{z=z_{15}}$.
They can be exposed if we make further $z$-shifts in momenta.
However, we do not indulge in these, because explicit expressions
for the four particle amplitudes are known already.

We turn to the residues at $z_{45}$.  The relevant twistors
for the cut line are
$$|\hat K_{45}\rangle=|4\rangle, \ \ \  |\hat K_{45}]=|4]+{\langle51\rangle\over \langle41\rangle}|5],$$
which form
$$\hat K_{45}\equiv (|4\rangle[4|+|\hat 5\rangle[5|)_{z=z_{45}}.$$
With these, after a fair amount of algebra, we
have
\bea
A_3^{(0)}(-\hat K_{45}{}^-,4{}^+,\hat 5{}^+)&
=& {[4 \hat 5]^3\over [4 \hat K_{45}][\hat K_{45}\hat 5]} \nonumber\\ &
=& {\langle14\rangle[45]\over \langle15\rangle},
\eea
\bea
A_3^{(1)}(-\hat K_{45}{}^+, 4{}^+,\hat 5{}^+)&
=&
-N{[4\hat 5][\hat 5\hat K_{45}][\hat K_{45} 4]\over K_{45}^2}\nonumber\\&
=& N{\langle15\rangle[45]^3\over \langle14 \rangle}{1\over K_{45}^2},
\eea
\bea
A_4^{(0)}(\hat 1{}^-,\hat 2^{}+,3{}^+,\hat K_{45}{}^-)&
=& {\langle\hat K_{45} \hat 1\rangle^3\over \langle\hat 1\hat 2\rangle\langle\hat 2 3\rangle\langle 3
\hat K_{45}\rangle}\nonumber\\ &
=& \frac{\langle 41\rangle^3}{\langle 12\rangle\langle 34\rangle}
\cdot \frac{[-2]\langle 41\rangle }{\langle 23\rangle [-2]\langle 41\rangle + \langle 45\rangle [5-]\langle 13\rangle},
\eea
and
\bea
 A_4^{(1)}(\hat 1{}^-,\hat 2{}^+,3{}^+,\hat K_{45}{}^+)
&=&N{\langle\hat 2 \hat K_{45}\rangle[\hat 2\hat K_{45}]^3\over
[\hat 1 \hat 2]\langle\hat 2  3 \rangle\langle 3 \hat K_{45}\rangle[\hat K_{45}\hat 1]}\nonumber\\ & =&
N\frac{(\langle 24\rangle [2-]+\langle 54\rangle [5-]) [23] \langle 13\rangle^3}{\langle 34\rangle^2}\frac{[-2]\langle 14\rangle }{(\langle 23\rangle [2-]\langle 41\rangle + \langle 45\rangle [5-]\langle 13\rangle)^2},\nonumber\\
\eea

From these, we form the partial amplitudes
\bea
A_{45}^{(a)}&\equiv & A_{4}^{(0)}(\hat 1{}^{-},\hat 2{}^{+}, 3{}^{+}, \hat K_{45}{}^{-}){1\over K_{45}^{2}}
A_{3}^{(1)}(-\hat K_{45}{}^{+}, 4{}^{+},\hat 5{}^{+})
\eea
and
\bea
A_{45}^{(b)}&\equiv & A_{4}^{(1)}(\hat 1{}^{-},\hat 2{}^{+}, 3{}^{+}, \hat K_{45}{}^{+}) {1\over K_{45}^{2}}
A_{3}^{(0)}(-\hat K_{45}{}^{-}, 4^{+},\hat 5^{+})
.
\eea

In the like manner, we construct the residues at $z=z_{23}$,
where the twistors are
$$|\hat K_{23}\rangle=|3\rangle, \ \ \ |\hat K_{23}]={\langle12\rangle\over \langle13\rangle}|2]+|3],$$
for
$$\hat K_{23} \equiv (|\hat 2\rangle[2|+|3\rangle[3|)_{z=z_{23}},$$
from which we obtain
\be
A_3^{(0)}(-\hat K_{23}{}^-,\hat 2{}^+,3{}^+)
={\langle13\rangle[23]\over \langle12\rangle},
\ee
\be
A_{3}^{(1)}(-\hat K_{23}{}^{+},\hat 2{}^{+},3{}^{+})
=-N{\langle12\rangle[23]^{3}\over \langle13\rangle}\frac{1}{K_{23}^{2}},
\ee
\bea
A_{4}^{(0)}(4{}^{+},\hat 5{}^{+},\hat 1{}^{-},\hat K_{23}{}^{-})
&=&\frac{\langle 1\hat K_{23}\rangle^3}{\langle \hat K_{23} 4\rangle \langle 4\hat 5\rangle \langle \hat 5 1\rangle },
\nonumber\\
&=&\frac{\langle 13\rangle^3 }{\langle 34\rangle\langle 51\rangle}\cdot
\frac{[-5]\langle 12\rangle}{\langle 23 \rangle [-2] \langle 41 \rangle + \langle 45 \rangle [5-]\langle 13\rangle}
\eea
and
\bea
A_4^{(1)}(\hat 1{}^-, \hat K_{23}{}^+, 4{}^+, \hat 5{}^+)
&=&N\frac{\langle \hat K_{23} \hat 5\rangle [\hat 5 \hat K_{23}]^3}{
[\hat 1\hat K_{23}]\langle \hat K_{23} 4\rangle\langle 4\hat 5\rangle
[\hat 5\hat 1]}\nonumber\\
&=&N\frac{(\langle 32\rangle [2-]+\langle 35\rangle [5-])[45]\langle 14\rangle^3}{\langle 34\rangle^2}\frac{[5-]\langle 13\rangle}{(\langle 23\rangle [2-]
\langle 41\rangle + \langle 45\rangle [5-]\langle 13\rangle)^2}.\nonumber\\
\eea
\bigskip
As before, we form partial amplitudes
\bea
A_{23}^{(a)}& \equiv&
 A_{4}^{(0)}(4{}^{+},\hat 5{}^{+},\hat 1{}^{-},\hat K_{23}{}^{-}){1\over K_{23}^{2}}
A_{3}^{(1)}(-\hat K_{23}{}^{+},\hat 2{}^{+},3{}^{+})
\eea
and
\bea
A_{23}^{(b)}& \equiv &
A_{4}^{(1)}( 4{}^{+},\hat 5{}^{+},\hat 1{}^{-},
\hat K_{23}{}^{+}){1\over K_{23}^{2}}
A_{3}^{(0)}(-\hat K_{23}{}^{-},\hat 2{}^{+}, 3{}^{+}).
\eea

In Figure 2 we depicted the remaining cut sub-amplitudes, $A_{23}^{a,b}$ and $A_{45}^{a,b}$. The conventions are the same as before. In addition, the letter $T$ now denotes a tree-level amplitude.

If we just add together $A_{12}+A_{15}+A_{23}^{(a)}+
A_{23}^{(b)}+A_{45}^{(a)}+A_{45}^{(b)}$, we shall include
the reducible contributions, which do not constitute the
complete amplitude $A_5$, as indicated by the fact that
this sum is dependent on the reference twistor $|-]$.  We need to add some
extra terms.  The central issue we posed for ourselves is
whether the demand of reference-twistor invariance of $A_5$ will
be sufficient to determine these extras.  We are going to
show the affirmative.  As we saw $A_{12}+A_{15}$ is $\eta$-independent.
Next, we should pair $A_{45}^{(a)}$
with $A_{23}^{(b)}$ and $A_{45}^{(b)}$ with $A_{23}^{(a)}$,
based on the appearance of factors
$\langle13\rangle^3[23]/\langle15\rangle$ and $\langle14\rangle^3[45]/\langle 12\rangle$
and others.  We propose  to add the extra terms separately
to these pairs and we further parameterize them as
$A_{45}^{(a)}f_{45}$ and $A_{23}^{(a)}f_{23}$.  We will show that
the expression of the 'soft factors` $f_{23}$ and $f_{45}$ can be secured by
requiring that
 $$A_{23}^{(b)}+ A_{45}^{(a)}(1+f_{45})$$
and $$A_{45}^{(b)}+ A_{23}^{(a)}(1+f_{23})$$ should
be made $\eta$-independent individually.

\begin{figure}[h!]
\begin{center}
$\begin{array}{c@{\hspace{.5 in}}c@{\hspace{.1in}}}
\includegraphics[width=2.5in]{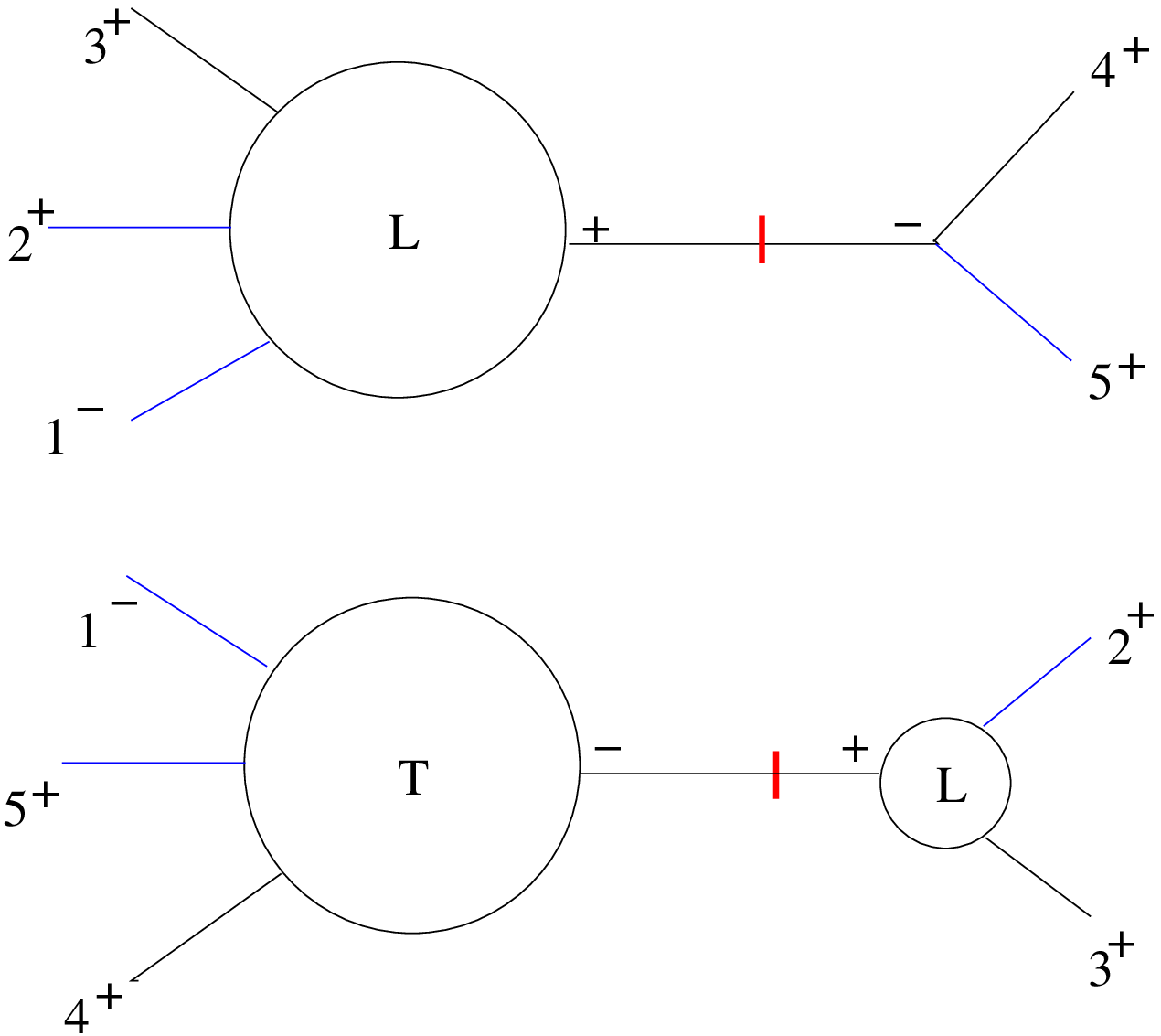}  &  
\includegraphics[width=2.5in]{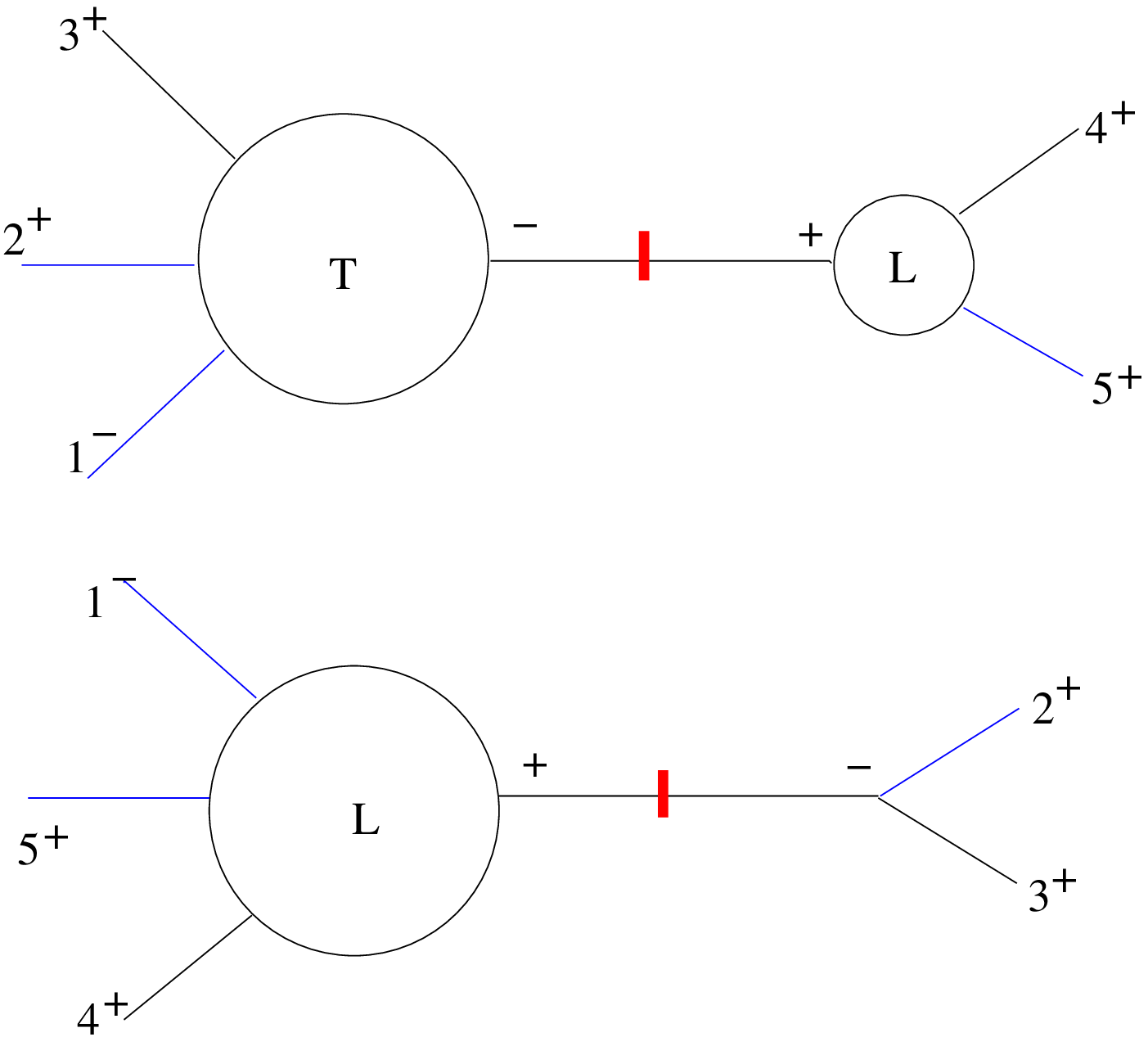}  \\
\end{array}$
\end{center}
\caption{$A_{45}^b$ and $A_{23}^a$ on the left side figure, and $A_{45}^a$ and $A_{23}^b$ to the right}
\end{figure}

The rationale behind the soft factors was explained in BDK. In factorizing the $(-++\dots+)$ amplitude into on-shell lower $n$-point functions, one encounters a double pole, corresponding to the one-part reducible graphs with a
$(P_i^+P_j^+K^+)$ one-loop vertex. The vertex itself is proportional with $\frac{1}{K^2}$, with an additional factor of $\frac{1}{K^2}$ coming from the propagator corresponding to the internal reducible line. However, there is an additional single pole which is underneath this double pole, and which is unmasked by the soft factor.
On the other hand, we can simply infer the necessity of adding the soft factor term contribution based on the knowledge that one-loop $(-++\dots+)$ amplitude is a rational function, and the observation that by accounting only for the double pole contribution we arrive at an expression which is not Lorentz covariant, as it depends explicitly on the reference twistor $|-]$. From this perspective, we can argue that the soft factor $f_{45}$ must be of the type
$$
f_{45}= \frac{A[2-]+B[5-]}{C[2-]+D[5-]}
$$
where $A,B,C,D$ depend only on the external momenta. Similar requirements hold for $f_{23}$.

The solution for $f_{45}$ is unique,
\be
f_{45}=-{\langle13\rangle\langle14\rangle\langle45\rangle(\langle32\rangle[2-]+\langle35\rangle[5-])\over
\langle15\rangle\langle34\rangle(\langle13\rangle\langle45\rangle[-5]+\langle14\rangle\langle23\rangle[2-])}.\label{(28)}
\ee
  Similarly, we find
\be
f_{23}=-{\langle13\rangle\langle14\rangle\langle23\rangle(\langle24\rangle[2-]+\langle45\rangle[-5])\over
\langle12\rangle\langle34\rangle(\langle13\rangle\langle45\rangle[-5]+\langle14\rangle\langle23\rangle[2-])}.\label{(30)}
\ee

In summary, we have shown that the one loop complete amplitude for $(1^-2^+3^+4^+5^+)$ is obtained recursively by
\be
A_5=A_{12}+A_{15}+A_{23}^{(a)}(1+f_{23})+
A_{23}^{(b)}+A_{45}^{(a)}(1+f_{45})+A_{45}^{(b)}.\label{5pt_Rec}
\ee

Since now we are guaranteed to arrive at $|-]$-independent expressions, we are free to choose the reference twistor $|-]$ such that the recursive relation is as simple as possible. There are two possible choices that lead to simplifications:
\bea
|-]=|2]\qquad  {\rm or}\qquad |-]=|5].
\eea
Then our triple shifts (\ref{(14)}) reduce to double shifts.
This is tantamount to using the BCFW analytic continuation, by singling out the momenta $P_1, P_2$ or $P_1, P_5$. Consequently, the soft factors (\ref{(28)}) and (\ref{(30)}) simplify to match precisely the expressions which were conjectured by BDK \cite{Bern:2005hs}, when using the double shift analytic continuation of the one-loop amplitude.

Lastly, let us compare (\ref{5pt_Rec}) with the known answer, obtained in \cite{Bern:1993mq}, using a string-inspired method   
\bea
A_5^{(1)}(1^-,2^+,3^+,4^+,5^+)&=& N{1\over \langle34\rangle^2}
\bigg[ -{[25]^3\over[12][51]}+{\langle14\rangle^3[45]\langle35\rangle\over \langle12\rangle\langle23\rangle\langle45\rangle^2} 
-
{\langle13\rangle^3[32]\langle42\rangle\over \langle15\rangle\langle54\rangle\langle32\rangle^2}\bigg].
\eea
We notice that the first term corresponds to the sum $A_{12}+A_{15}$. The second and third term correspond precisely to the sums $A_{23}^{(b)}+ A_{45}^{(a)}(1+f_{45})$ and $A_{45}^{(b)}+ A_{23}^{(a)}(1+f_{23})$.

\subsubsection{The Six-Point $(-+++++)$ Amplitude}

In order to discern a general pattern for the on-shell recurrence relation, we work out explicitly  
the next $(-++\dots +)$ one-loop amplitude, namely the six-point $(1^-2^+3^+4^+5^+6^+)$.

We perform the  analytic continuation to $A(z)$ by means of the triple shift
 \bea
&&[\hat 1|=[1|+z[26][-|\nonumber\\
&&|\hat 2\rangle=|2\rangle+z[6-]|1\rangle\nonumber\\
&&|\hat 6\rangle=|6\rangle+z[-2]|1\rangle. \label{(126)}
\eea
The terms in the recursive relation correspond to poles in $\hat K_{16}^2, \hat K_{12}^2, \hat K_{23}^2,\hat K_{56}^2,\hat K_{234}^2, \hat K_{456}^2$. The terms which factorize into a loop $(+++)$ vertex are obtained, again, from a double pole. These terms are the ones that need a soft-factor correction in order to arrive at an expression that is reference-twistor independent.
Along the way we will find out which are the combinations of cut graphs that are separately $\eta$-independent.

The first observation is that the sum $A_{12} + A_{16}$, where
\bea
A_{12}& \equiv& A_{3}^{(0)}(\hat 1{}^{-},\hat 2{}^{+},-\hat K_{12}{}^{-})|_{z=z_{12}}
{1\over K_{12}^{2}}A_{5}^{(1)}(\hat K_{12}{}^{+} ,\hat 3{}^{+},4{}^{+},5{}^{+},\hat 6^+)|_{z=z_{12}}\nonumber\\
A_{16}& \equiv& A_{3}^{(0)}(\hat 6{}^{+},\hat 1{}^{-},-\hat K_{16}{}^{-})|_{z=z_{16}}
{1\over K_{16}^{2}}A_{5}^{(1)}(\hat K_{16}{}^{+} ,\hat 2{}^{+},3{}^{+},4{}^{+},5^+)|_{z=z_{16}}
\eea
is $\eta$-independent:
\bea
A_{12}+A_{16}=N\frac{[26]^3}{[12][61]s_{345}}\bigg(\frac{[23][34]}{\langle 45\rangle\langle 5|3\!\!\slash{} +4\!\!\slash{}|2]}-\frac{[45][56]}{\langle 34\rangle\langle 3|1\!\!\slash{}+2\!\!\slash{}|6]}+\frac{[35]}{\langle 34\rangle\langle 45\rangle}\bigg).
\label{12+16}
\eea
\begin{figure}[!h]
\begin{center}
\includegraphics[height=3in]{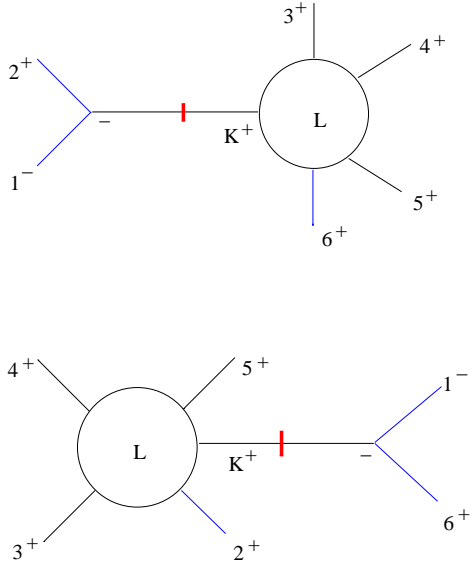}
\caption{A group of cut sub-amplitudes, $A_{12}$ and $A_{16}$ which involve a five-point $(+++++)$ factor. Their sum is reference-twistor independent and given in (\ref{12+16}).}
\end{center}
\end{figure}

The terms which arise from the $\hat K_{234}^2$ and $\hat K_{456}^2$ poles, after some algebra, can be written as:
\bea
A_{234}&\equiv& A_4^{(0)}(\hat 1^-,\hat K_{234}^-,5^+,6^+)\frac{1}{K_{234}^2}
A_4^{(1)}(-\hat K_{234}^+, \hat 2^+, 3^+,4^+)\nonumber\\
&=&-N\frac{\langle 1|3\!\!\slash+4\!\!\slash|2]^3}{\langle 5|3\!\!\slash+4\!\!\slash|2]\langle 16\rangle\langle 34\rangle^2}\frac{1}{s_{234}}\frac{1}{\langle5\hat 6_{234}\rangle}\\
A_{456}&\equiv&A_4^{(0)}(\hat 1^-,2^+,3^+,\hat K_{234}^-)\frac{1}{K_{456}^2}
A_4^{(1)}(-\hat K_{234}^+, 4^+, 5^+,\hat 6^+)\nonumber\\
&=&-N\frac{\langle 1|4\!\!\slash+5\!\!\slash|6]^3}{\langle 12\rangle \langle 45\rangle^2 \langle 3|4\!\!\slash+5\!\!\slash|6]}\frac{1}{s_{456}}
\frac{1}{\langle 3\hat 2_{456}\rangle},
\eea
where the twistor products $\langle \hat 2_{456} 3\rangle $ and $\langle 5 \hat 6_{234}\rangle$ are
\bea
\langle \hat 2_{456}3\rangle =\langle 23\rangle + \frac{s_{456}[6-]\langle 31\rangle}{[2-]\langle 1|4\!\!\slash + 5\!\!\slash|6]}
\\
\langle 5\hat 6_{234}\rangle =\langle 56\rangle +\frac{s_{234}[-2]\langle 51\rangle}{[6-]\langle 1|3\!\!\slash + 4\!\!\slash|2]}.
\eea
These terms do not combine in any way, and their sum remains reference-twistor dependent.

Moving on, we have also terms which arise from the factorization onto
$\hat K_{23}^2$ and $\hat K_{56}^2$:
\bea
A_{23}^b=A_5^{(1)}(-\hat K_{23}^+, 4^+,5^+,\hat 6^+, \hat 1^-)\frac{1}{K_{23}^2}
A_3^{(0)}(\hat K_{23}^-,\hat 2^+, 3^+)\\
A_{56}^b=A_5^{(1)}(-\hat K_{56}^+, \hat 1^-,2^+,3^+,4^+)\frac{1}{K_{56}^2}
A_3^{(0)}(\hat K_{56}^-, 5^+,\hat 6^+).
\eea
Let us focus of $A_{23}^b$.
We saw in the previous section that the one-loop five--point functions can be written as a sum of three terms. Substituting into  $A_{23}^b$, we arrive at:
\bea
A_{23}^b=\frac{N}{\langle 45\rangle^2}\bigg(-\frac{[\hat K_{23} 6]^3}
{[\hat 1 \hat K_{23}][ 6\hat 1]}+\frac{\langle 15\rangle^3[56]\langle4\hat 6\rangle}{\langle 1 \hat K_{23}\rangle \langle \hat K_{23} 4\rangle \langle 5\hat 6\rangle^2}
+\frac{\langle 14\rangle^3 [4 \hat K_{23}]\langle 5 \hat K_{23}\rangle }{\langle 16\rangle\langle \hat 6 5\rangle \langle 4 \hat K_{23}\rangle^2}\bigg)\frac{1}{\langle23\rangle[23]}\frac{[23]^3}{[2\hat K_{23}][\hat K_{23} 3]},\nonumber\\
\eea
where
\bea
z_{23}=-\frac{\langle 23\rangle}{[6-]\langle 13\rangle},\qquad
|\hat K_{23}\rangle=-|3\rangle, \qquad |\hat K_{23}]= |3]+\frac{\langle 12\rangle}{
\langle 13\rangle}|2].
\eea
After a bit of massaging, the first term in $A_{23}^b$ can be cast into
\be
N\frac{\langle 1|4\!\!\slash + 5\!\!\slash|6]^2 \langle 13\rangle [6-]}{
\langle 45\rangle^2 [2-]\langle 12\rangle \langle 23\rangle \langle 3\hat 2_{456}\rangle
\langle 3|4\!\!\slash + 5\!\!\slash|6]}.
\ee
This combines with $A_{456}$ into an $\eta$-independent expression:
\be
N\frac{\langle 1|2\!\!\slash+3\!\!\slash|6]^3}{\langle 12\rangle\langle 23\rangle\langle 45\rangle^2 s_{123}\langle 3|1\!\!\slash+2\!\!\slash|6]}.\label{6pta}
\ee
Similarly, $A_{234}$ combines with one of the three terms on $A_{56}$ into
an expression that is manifestly reference-twistor independent
\be
N \frac{\langle 1|3\!\!\slash+4\!\!\slash|2]^3}{\langle 34\rangle^2\langle 56\rangle\langle 61\rangle s_{234}\langle 5|3\!\!\slash+4\!\!\slash|2]}.\label{6ptb}
\ee
\begin{figure}[!h]
\begin{center}
$\begin{array}{c@{\hspace{.4 in}}c@{\hspace{.1in}}}
\includegraphics[width=3.2in]{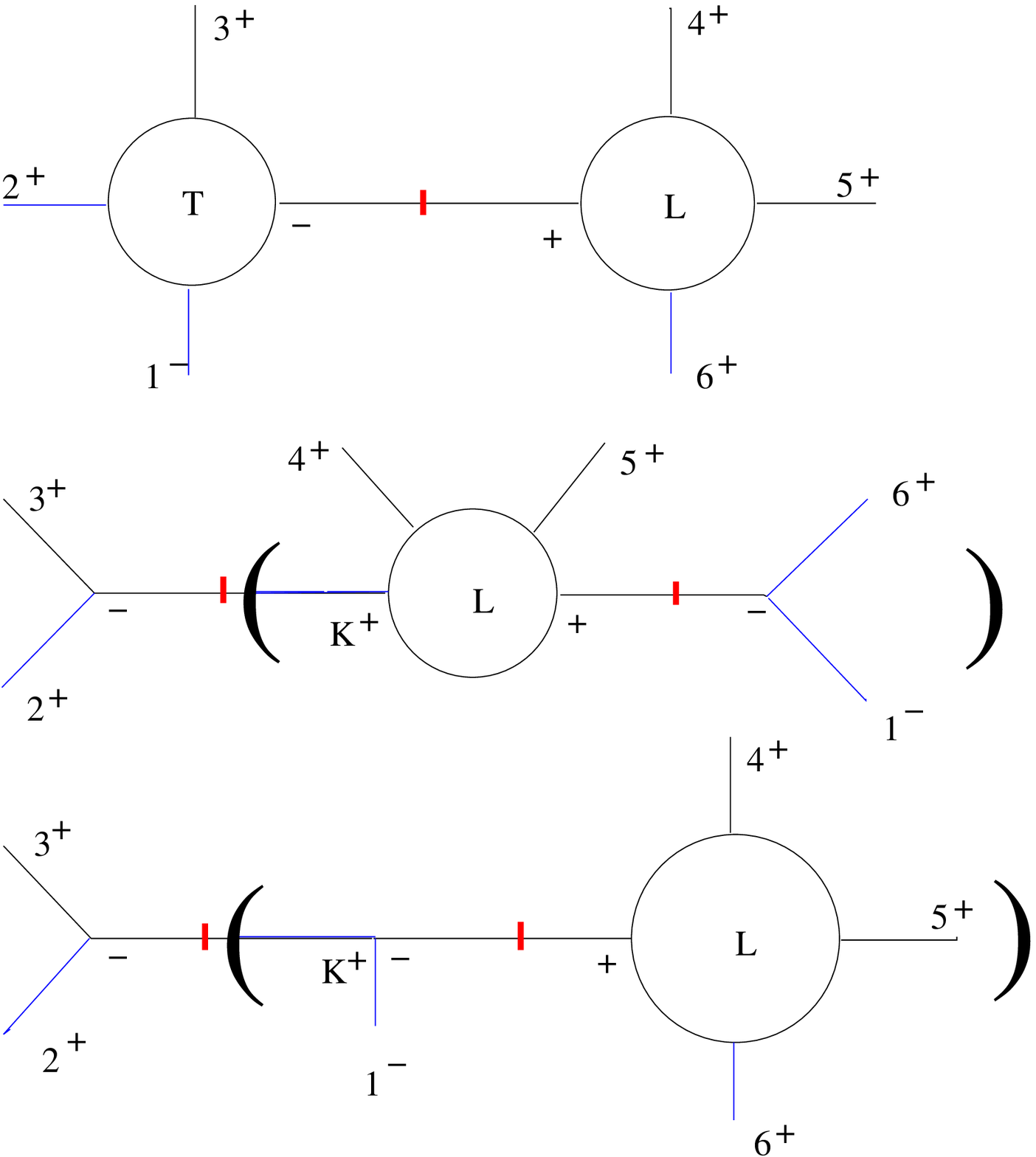}  &  
\includegraphics[width=3.2in]{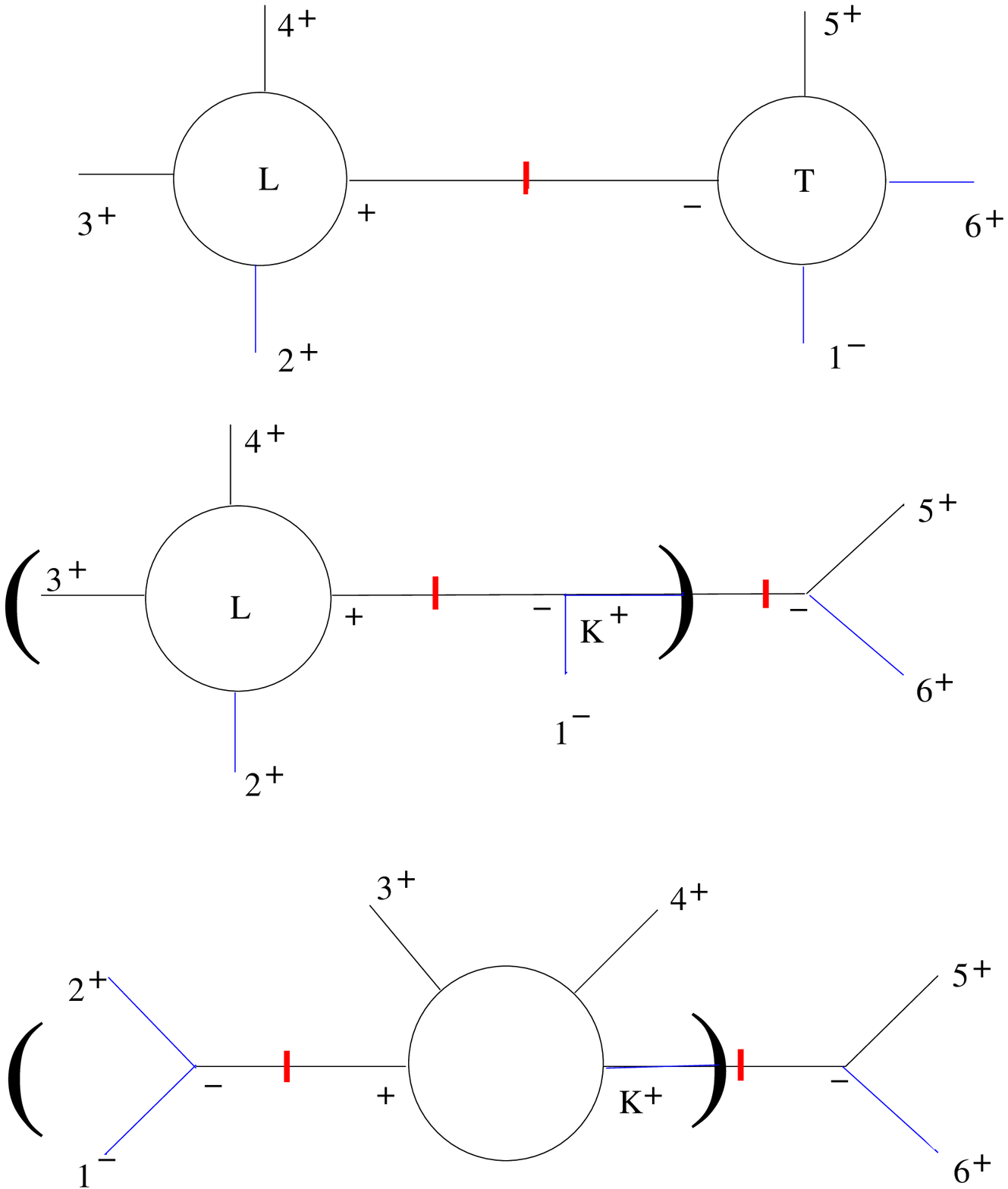}  \\
\end{array}$
\end{center}
\caption{ These cut sub-amplitudes contain a lone-loop (++++) on-shell factor. Their sum is reference-twistor invariant (separately for the left and right cut sub-amplitudes), corresponding to (\ref{6pta}) and (\ref{6ptb}).}
\end{figure}

In Figure 4 we have depicted the cut sub-amplitudes which add up to (\ref{6pta}) and respectively (\ref{6ptb}). The novel feature is the presence of a nested double shift: first one places $K_{23}$ (or $K_{56}$) on-shell via the $P_1,P_2,P_6$ shift; this would factorize the six-point amplitude into a tree-level times a one-loop five-point $(-++++)$ amplitude. As explained before, we have to fish out certain terms out of the one-loop five-point amplitude in order to expose the terms which form $\eta$-independent expressions. These terms are obtained by cuts following from a second, subsequent shift, involving again, the lone negative helicity gluon $1^-$ and the two adjacent on-shell positive helicity gluons. This subsequent cut is performed in the part of the diagram in between the big bold brackets, and the line which is placed on-shell is shown by the same token of a small (red) vertical cut.

One term each from $A_{23}^b$ and $A_{56}^b$, those proportional to $\langle 14\rangle^3$, add up to another $\eta$-independent combination:
\be
-\frac{\langle 14\rangle^3 \langle 35\rangle \langle 1|2\!\!\slash+3\!\!\slash|4]}{\langle 12\rangle\langle 23\rangle\langle 34\rangle^2\langle 45\rangle^2\langle 56\rangle\langle 61\rangle}\label{6pt14}.
\ee

\begin{figure}[!h]
\begin{center}
\includegraphics[height=4.4in]{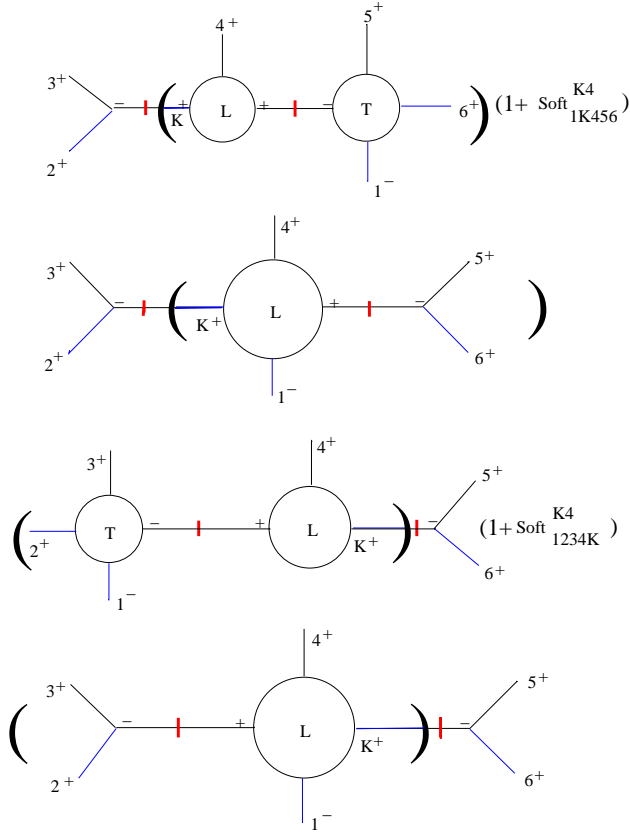}
\caption{A group of cut sub-amplitudes which adds up to a reference-twistor invariant expression, (\ref{6pt14}).}
\end{center}
\end{figure}

At this moment we are left with one term each from $A_{23}^b$ and $A_{56}^b$, specifically those proportional to $\langle 15\rangle^3[56]$ and $\langle 13\rangle^3[23]$ respectively, and the terms which arise from the factorization onto
the double poles $(\hat K_{23}^2)^2$ and $(\hat K_{16}^2)^2$:
\bea
A_{23}^a &\equiv& A_5^{(0)}( 4^+,5^+,\hat 6^+,\hat 1^-,\hat K_{23}{}^-)
\frac{1}{K_{23}^2}A_3^{(1)}(-\hat K_{23}{}^+,\hat 2^+,3^+)\\
A_{56}^a&\equiv& A_5^{(0)}(\hat 1^-,\hat 2^+,3^+,4^+,\hat K_{56}{}^-)
\frac{1}{K_{56}^2}A_3^{(1)}(\hat K_{56}{}^+,5^+,\hat 6^+).
\eea
To arrive at the known result, BDK \cite{Bern:2005hs} conjectured the existence of soft-factor terms, whose role is to account for the unknown single-pole-underneath-the-double-pole contribution, in evaluating $\oint \frac {dz}{ z} A(z)$. However, from our perspective, we simply observe that the double-pole terms, without any additional contribution, together with the remaining terms from $A_{23}^b$ and $A_{56}^b$, add up to something which is not Lorentz covariant, as it contains an explicit dependence on the reference twistor $|-]$. We are led to the conclusion that an additional contribution is needed to cure this problem. We require that the missing terms satisfy
\bea
\frac{\partial}{\partial |-]}\bigg( A_{23}^b\bigg|_{{\rm keep\;only\;the\;}\langle 15\rangle^3-{\rm term}}+A_{56}^a (1+ f_{56})\bigg)=0
\eea
and similarly,
\bea
\frac{\partial}{\partial |-]} \bigg(A_{56}^b\bigg|_{{\rm keep\;only\;the\;}\langle 13\rangle^3-{\rm term}}+A_{23}^a (1+ f_{23})\bigg)=0,
\eea
where $f_{23}$ and $f_{56}$ are of the type
\bea
\frac{A[2-]+B[6-]}{C[2-]+D[6-]}.\nonumber
\eea
It turns out that the soft factors are completely determined by the reference-twistor independence requirement.
 We compute
\bea
f_{23}&=&-\frac{\langle 23\rangle \langle 13\rangle(\langle 24\rangle[2-]\langle 15\rangle+\langle 65\rangle[6-]\langle 14\rangle )}{\langle 12\rangle\langle 34\rangle(\langle 56\rangle [6-]\langle 13\rangle+\langle 32\rangle[2-]\langle 15\rangle)} \\
f_{56}&=&\frac{\langle 56\rangle\langle 15\rangle(\langle 64\rangle[6-]\langle 13\rangle+\langle 23\rangle[2-]\langle 14\rangle)}{\langle 16\rangle\langle 45\rangle(\langle 56\rangle [6-]\langle 13\rangle+\langle 32\rangle[2-]\langle 15\rangle)}.
\eea
At this stage we can again revert to making the identification $|-]=|2]$ or,
$|-]=|6]$. Then, the triple momentum shifts reduce to standard (BCFW) shifts, and the soft factors determined previously reduce to the expressions which have been conjectured by BDK \cite{Bern:2005hs}.
\begin{figure}[!h]
\begin{center}
$\begin{array}{c@{\hspace{.4 in}}c@{\hspace{.1in}}}
\includegraphics[width=3.2in]{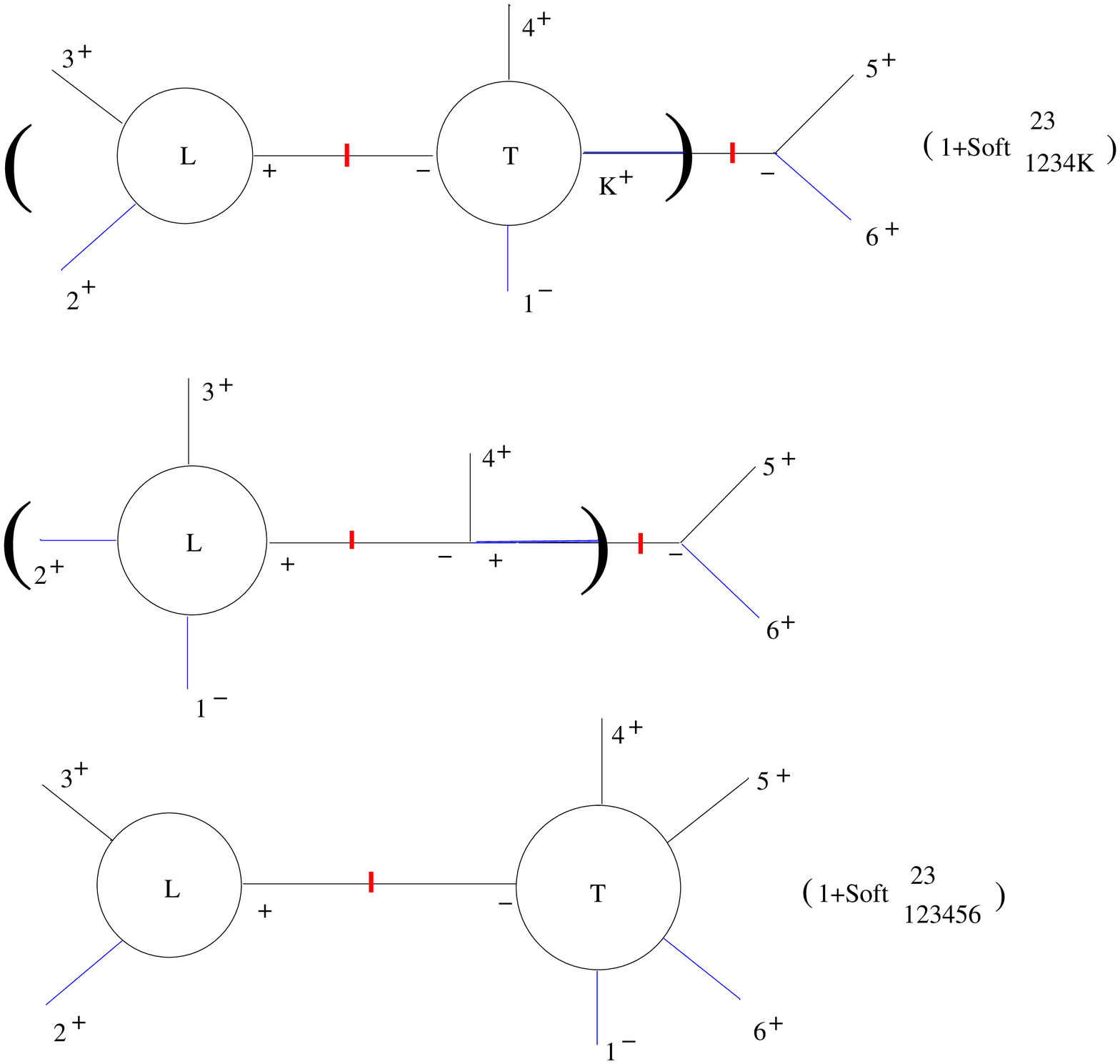}  &  
\includegraphics[width=3.2in]{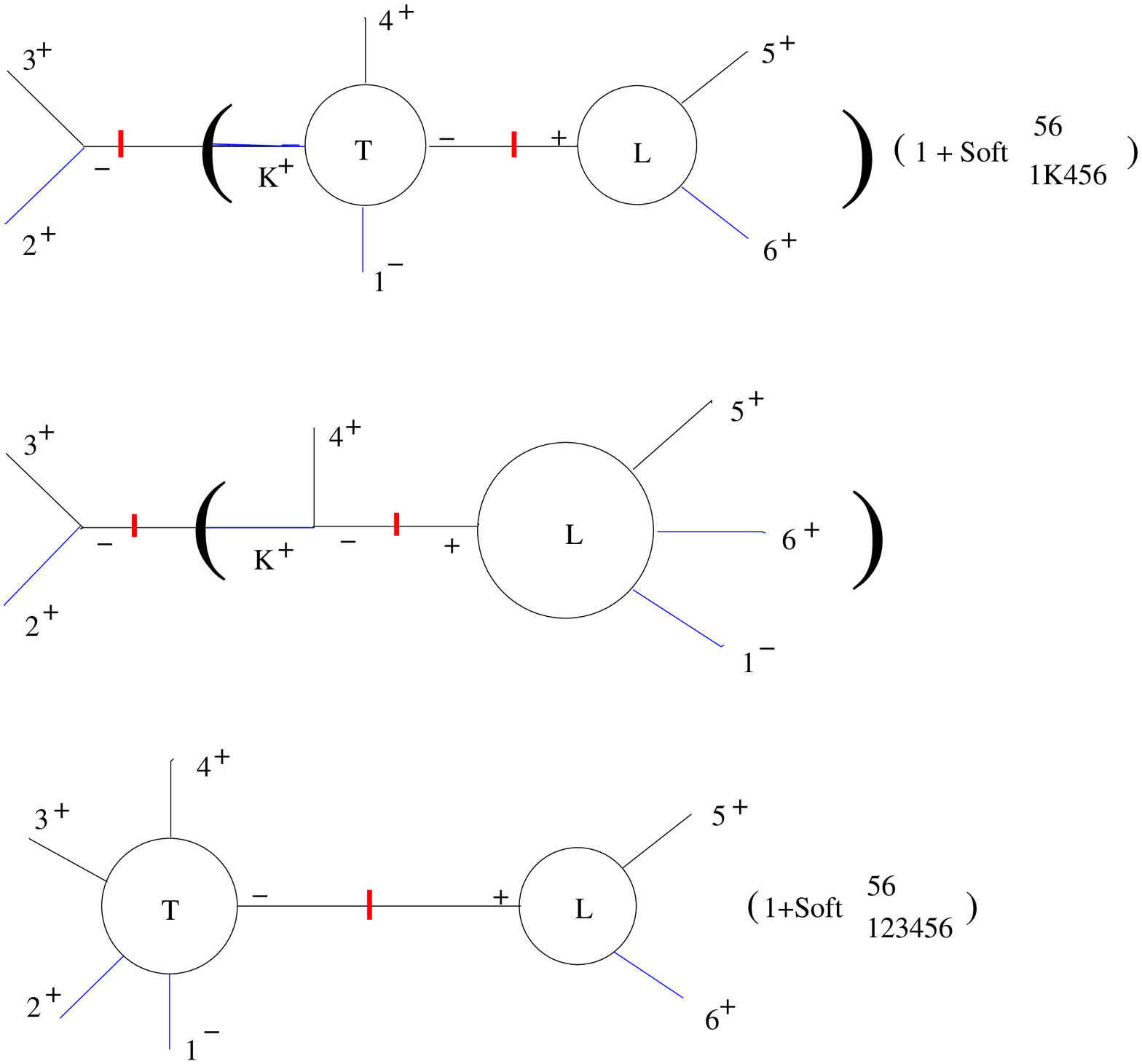}  \\
\end{array}$
\end{center}
\caption{ The cut sub-amplitudes on the left and on the right add up separately to reference-twistor invariant expressions. This is used to determine the six-point function soft factors.}
\end{figure}

\subsubsection{A generic (-++\dots +) Amplitude }

From the results above, we can infer that for a
general amplitude $(1^-2^+\dots n^+)$, the soft
factors are
\bea
f_{23}=\big({\langle13\rangle\langle23\rangle\over \langle12\rangle\langle34\rangle}\big)
\big( {[n\ -]\langle 14\rangle\langle n-1\  \rangle-[-2]\langle24\rangle\langle n-1\  1\rangle\over
[n\ -]\langle 13\rangle \langle n-1\  n\rangle -[-2]\langle 23\rangle \langle n-1\  1\rangle }\big),
\eea
and
\bea
f_{n-1 \ n}&=&\big({\langle 1\ n-1\rangle \langle n\ n-1\rangle \over \langle 1\ n\rangle \langle n-1\ n-2\rangle }\big)
\nonumber\\
&\times&
\big({[n-]\langle 13\rangle\langle n-2\ n\rangle -[-2]\langle 23\rangle \langle n-2\ 1\rangle  \over
[n\ -]\langle 13\rangle \langle n-1\  n\rangle -[-2]\langle 23\rangle \langle n-1\  1\rangle }\big),
\eea
where we have made the shifts given in (\ref{gen_shifts}).

To gain some understanding, let us make some comments.
Because of the symmetry of the problem, the amplitude
satisfies a reflection property
\be
A(1^-2^+3^+\dots n^+)=(-1)^n A(1^- n^+ (n-1)^+\dots 2^+),
\ee
which implies that
\be
f_{23}(12\dots n)=f_{n-1\ n}(1 \ n \ (n-1)\dots 2).
\ee
This symmetry under $2 \to n, \ 3\to n-1 \dots n\to 2$
is maintained in our shifts, as long as $|-]$ is kept
general.  Thus to evolve the soft factors from
\be
|-] \to |-]+|\delta],
\ee
where $|\delta]$ is some infinitesimal twistor,
they must have the same rate
\be
\delta f_{23}(1 2\dots n)=-\delta f_{n-1\ n}(12\dots n).
\ee
A little calculation gives
\be
\delta  f_{23}=-{z_{23}z_{n-1\ n}\over (z_{23}-z_{n-1\ n})^2}
{[2n]\over [-\ 2] [n\ -]}([- \delta]),
\ee
where
\be
z_{23}=-{\langle 23\rangle \over [n\ -]\langle 13\rangle }, \ \ \ \
z_{n-1 \ n}=-{\langle n-1 \ n\rangle \over [2\ -]\langle 1\ n-1\rangle },
\ee
are respectively the poles in the $z$-plane corresponding to the
$(23)$ and $(n\ n-1)$ channels.  We can understand
the factor $z_{23}z_{n-1\ n}$ in the numerator,
because under evolution  of $f_{23}$ and $f_{n-1\ n}$
from some initial values
to the value $|-]$ we want, they will provide a factor
$\langle 23\rangle $ and $\langle n-1\ n\rangle $, respectively, to convert
a double pole into a simple pole in the partial
amplitudes of those channels.  The other factors are there to make
$\delta f_{23}$ and $\delta f_{n-1\ n}$ into something
which depends only on the pole structure and
the choice of momenta (2 and $n$) which are shifted.
A deeper reason for this is needed, however.

In this respect, we want to point out that one can obtain
the soft factors from just a single term for each amplitude.  
Thus, for $f_{23}$ of the five-point amplitude, it is the ratio of the simple
pole over the double pole of $z_{23}$ of
$$
N{\langle 13\rangle^3[23]\langle 24\rangle \over \langle 23\rangle^2\langle 34\rangle^2}{1\over\langle 45\rangle\langle 51\rangle};
$$
for the six-point amplitude, it is
$$
N{\langle 13\rangle ^3[23]\langle 24\rangle \over \langle 23\rangle^2\langle 34\rangle ^2}{1\over \langle 45\rangle \langle 56\rangle \langle 61\rangle};
$$
and for the seven-point function, it is
$$N{\langle13\rangle^3[23]\langle24\rangle\over \langle 23\rangle^2\langle 34\rangle ^2}
{1\over \langle 45\rangle \langle 56\rangle \langle 67\rangle \langle 71\rangle}.
$$
From the perspective of grouping the cut sub-amplitudes, where successive cuts have been performed until a loop $(++\dots+)$ has been exposed, and with the cuts implemented by shifts always involving the lone negative helicity gluon $1^-$  and its adjacent on-shell positive helicity gluons (as in Figures 4,5,6), the terms which add up to separately $\eta$-independent expressions are those where the loop $(++\dots+)$ amplitude is a four-point or higher amplitude. The terms which need a soft factor improvement are those where the cuts have revealed a $(2^+3^+K^+)$ (or $K^+ \ (n-1)^+ \ n^+)$) amplitude and a one-loop $(1^-+++)$ amplitude, with one of the positive helicity gluons being either $2^+$ or $n^+$.

Since Bern, Dixon and Kosower \cite{Bern:1993mq} obtained their results for the one-loop five-point function $(-++++)$ using a string-inspired method \cite{Bern:1991aq},
it seems that a natural setting is to analytically continue
their approach to complex momenta in the $z$-plane
and study its structure there, as imposed by the
requirement of cut
reference independence of the physical result at $z=0$.

\section{Concluding Remarks}

In past discussions of helicity twistor evaluation of
QCD amplitudes, analyticity, complex momenta and
unitarity are the main topics and their deployments
have led to an impressive list of accomplishments.
We have emphasized and shown in this article that
the freedom of gauge choice and the requirement
of Lorentz invariance for physical amplitudes
make for another powerful tool, which should be exploited.
As a concrete case, we have used it to derive recursion
relations and to determine the soft factors in
(-++\dots +) amplitudes.  One noteworthy feature
is that the evolution of these soft factors is completely
given by the pole structure in the $z$-plane and our
choice of shifted momenta and reference twistors
$|+\rangle, \ [-|.$  Hence, gauge freedom and analyticity
find their meeting place.  We are hopeful that their
interplay and mutual reinforcement will help us understand
the structural properties of amplitudes with other
helicity arrangements and  construct them.

We have made extensive use of the freedom in
the space-cone gauge to dictate our choice of shifted
momenta.  Because the asymptotic behavior
$A(z\to \infty) \to 0$ is automatically satisfied as a
result of vertex compositions, recursion relations
follow without complication from Cauchy theorem
for (++\dots +) and (-++\dots +) amplitudes.  We
have also used holomorphic shifts to obtain recursion
relations for the former.

\section*{Acknowledgments}
YPY would like to thank Professor C. N. Yang and Professor H. T. Nieh
for hospitality at the Center for Advanced Studies, Tsinghua University.

This research is supported in part by the National Science Foundation under grant PHY05-61164.

\end{document}